\documentclass{cernrep}
\pdfoutput=1
\usepackage{tikzfeynman}		
\usetikzlibrary{backgrounds}
\usetikzlibrary{mindmap,trees}
\usepackage{texnames}
\usepackage[T1]{fontenc}
\usepackage[bookmarks, colorlinks=true, linktoc=page, pdftex, linkcolor=red, citecolor=red, urlcolor=red]{hyperref}
\usepackage{pdfcomment}
\begin{document}
\title{Heavy-ion physics: freedom to do hot, dense, exciting QCD}
 
\author{Mar\'ia Elena Tejeda-Yeomans}

\institute{Facultad de Ciencias - CUICBAS\\ Universidad de Colima \\ M\'exico}

\begin{abstract}
In these two lectures I review the basics
of heavy-ion collisions at relativistic energies and the physics we can do with them. I aim to cover the basics on the kinematics and observables in heavy-ion collider experiments, the basics on the phenomenology of the nuclear matter phase diagram, some of the model building and simulations currently used in the heavy-ion physics community and a selected list of amazing phenomenological discoveries and predictions.
\end{abstract}

\keywords{CLASHEP, CERN; QCD; heavy-ion; quark-gluon plasma; QCD phase diagram; temperature; baryon density.}

\maketitle 
 
\section{The not-so-simple questions that remain.}

The recent discoveries in particle physics are built upon the platform created by a solid quantum field theory of the strong interactions: Quantum Chromodynamics (QCD). Without it, we could not "see" through the haze of multiple particle production at the LHC and study different decay channels of the Higgs boson. Eventhough QCD is robust enough to encompass perturbative and non-perturbative approaches that allow us to study a variety of phenomena in hadron colliders, we are always tempted to put it to the test in the high temperature and/or density regimes. For example, if we already know how to study proton-proton collisions ($p+p$ collisions) at LHC energies, starting with parton level cross sections using QCD matrix elements that involve quarks and gluons, that later become hadron level cross sections and can be connected to shape obervables or jet observables, how much of this knowledge can be used to simulate jet formation in nucleus-nucleus collisions ($A+A$ collisions, where $A$ can be any nucleus)?. Do we assume that hadronization and jet formation in the collision of large systems happens just as it does in small systems? It turns out that these questions are just the tip of the iceberg: the physics of heavy-ion collisions at relativistic energies are a portal to new phenomena of the strong interactions. They are the means to explore regimes of QCD under \textit{extreme} conditions: what happens to strongly interacting matter when it is in a thermal bath and/or when it is highly compressed? does this mean that we expect ordinary nuclear matter to phase transition into new states of matter? how do we know theoretically and experimentally what are the signals we should look for when these phase transitions occur? Suddenly, heavy-ion collision experiments represent our best tool to learn about the questions that remain: the rich dynamical QCD phenomena that emerges under these conditions, an arena to further extend the vast knowledge we already have on hadron and QCD physics.

Now, if you are already an expert on using QCD to study $p+p$ collisions, then you might be surprised to learn that from there, it is an easy road to learn more about $A+A$ collisions. This is the road I plan to take in these couple of lectures, to discuss about the phases of nuclear matter, about QCD now under extreme conditions, about quantum field theory techniques for larger systems and about code used in $p+p$ collisions to be able to simulate $A+A$ collisions. In these two lectures I will aim to give you the basics of heavy-ion collisions at relativistic energies and the physics we can do with them. These lectures will rely heavily on previous enlightening talks and short courses given at different venues (see for example \cite{Talks2000}) and on various articles and textbooks~\cite{Busza2018, Wong1994, Csernai1994, Shuryak2004, Vogt2007, Flork2010, Sarkar2010}.

The structure of the lectures is as follows: in Section~\ref{sec:QCDu} we review the essential ideas that helped establish QCD as the theory of strong interactions and use them to have an initial discussion about the nuclear matter phase diagram, then in Section~\ref{sec:rhics} we review the basic ideas used to study relativistic heavy-ion collisions and the evolution of the system after the collisions, in Section~\ref{sec:theqgp} we discuss the quark-gluon plasma and its basic properties, in Section~\ref{sec:phasecritical} we introduce the tools required to study the phase diagram of nuclear matter and the observables with which phase transitions and critical behavior can be probed, finally in Section~\ref{sec:seasons} we highlight observables that help characterize the QGP formation and evolution. We provide final remarks and future prospects in Section~\ref{sec:final}.

\section{- \textit{QCD as usual} - said no one, ever.}
\label{sec:QCDu}

\subsection{The basics}

\begin{figure}
\centering\includegraphics[scale=0.6]{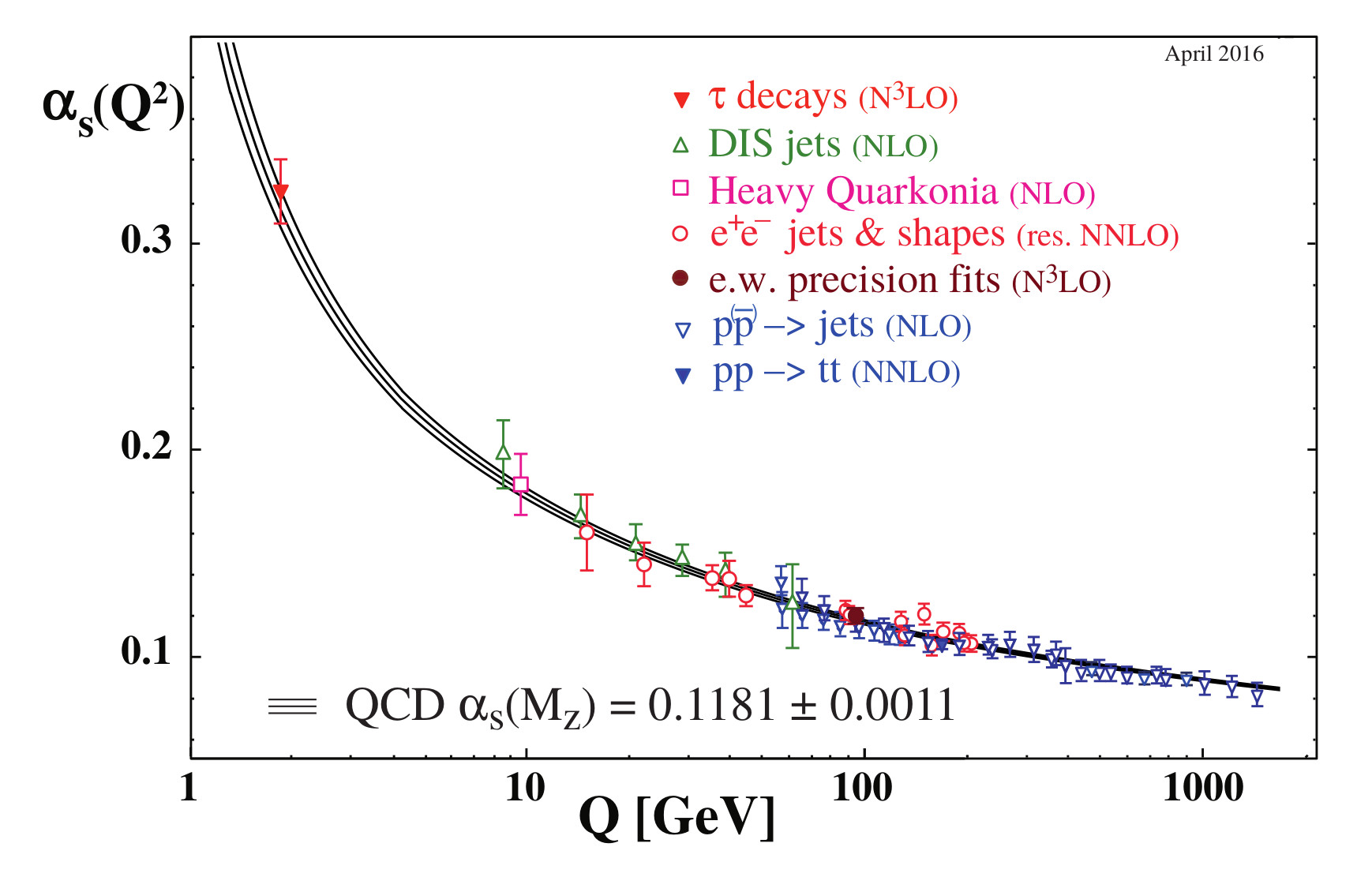}
\caption{Measurements of $\alpha_s$ as a function of the energy scale $Q$, as reported in the Quantum Chromodynamics review of the Particle Data Group ~\cite{PDG2018}. From each set of experiments, there is a value of $\alpha_s$ that has been extracted using different levels of QCD perturbation theory, which are indicated in brackets.}
\label{fig:asq2016}
\end{figure}

It is now carved in stone (which is to say, it has been established by numerous experiments) that nature favours an $SU(3)_c$ gauge theory as the mathematical framework with huge phenomenological consequences that shape our current understanding of the strong interactions at a fundamental level. As is shown in Fig.~\ref{fig:asq2016}, QCD, the theory of strong interactions that Nature favours, is everywhere. The emblematic characteristics of this theory make it unique and powerfull: $\alpha_S$ running with the energy scale of a wide range of experiments and $SU(3)_c$ well established as the underlying symmetry\footnote{For more interesting discussions about the experimental tests of QCD and about how to extract $\alpha_s$ beyond perturbative QCD, see for example~\cite{QCDstatus, stan}.}. This supports the formulation of a microscopic theory in terms of elementary fields (quarks and gluons), whose interactions obey the principles of a relativistic quantum field theory with non-abelian gauge invariance.

\begin{figure}
\begin{center}
		\begin{tikzpicture}[scale=0.7,transform shape]
			\Huge
			\begin{scope}[mindmap, concept color=red, text=white, scale=0.7,transform shape]
			\tikzstyle{level 1 concept}+=[sibling angle=120]
			\tikzstyle{level 2 concept}+=[sibling angle=45]
			
			\tikzstyle{root concept}+=[minimum size=1.5cm, text width=1cm] 
			
			\node[concept] (qcd) {QCD}
				[clockwise from = 35]
				child[concept color=blue]{node[concept] (npqcd) {npQCD}
					[clockwise from = 45]
					child[concept color=blue]{node[concept] (lattice) {lattice}}
					child[concept color=blue]{node[concept] (sde) {SDEs}}
					child[concept color=blue]{node[concept] (sumr) {sum rules}}
				}
				child[concept color=green!50!black]{node[concept] (effective) {effective theories}
				    [clockwise from = -45]
					child[concept color=green!50!black]{node[concept] (hqet) {heavy quark ET}}
					child[concept color=green!50!black]{node[concept] (scet) {SCET}}
					child[concept color=green!50!black]{node[concept] (nrqcd) {nrQCD}}
				}
				child[concept color=black!50!red]{node[concept] (pqcd) {pQCD}
                    [clockwise from = -140]
					child[concept color=black!50!red]{node[concept] (mcs) {loops, legs $\to$ $\epsilon$ structure}}
					child[concept color=black!50!red]{node[concept] (beyond) {duality}}
					child[concept color=black!50!red]{node[concept] (factor) {factorization, jets $\to$ MCs}}
				}
			;
			\end{scope}
		\end{tikzpicture}
	\end{center}
\caption{QCD is everywhere. Non-perturbative QCD (npQCD) has a vast reach and we here highlight Schwinger-Dyson equations (SDEs) are solved for the quark and gluon propagator with dressed versions of the QCD vertices to get hadron masses. Lattice QCD can solve numerically the QCD equations of motion and produce incredible results for hadron spectra and thermodynamic properties of nuclear matter. Perturbative QCD (pQCD) was the historical way to establish QCD as a gauge theory of the strong interactions and provided the tools to predict and describe \textit{jets}, which are now known as the standard signature of hard processes. Throughout the development of hadron and electroweak physics, effective theories have paved the way for modern QCD phenomenology.}
\label{fig:qcdish}
\end{figure}
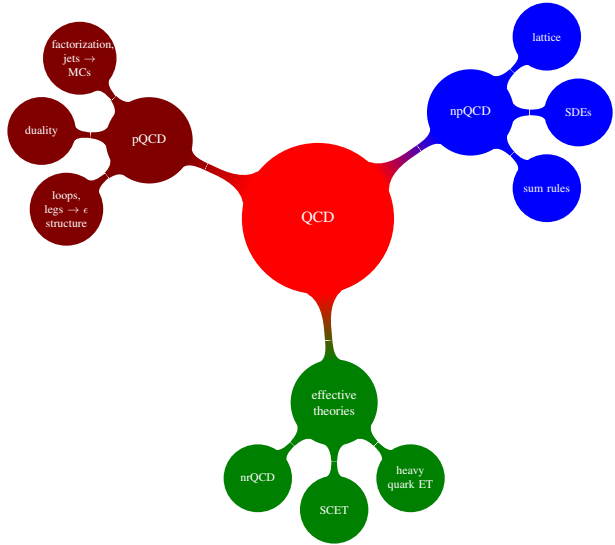

Since the early development of QCD, people started to amass tools that would allow to study pheonomena at the heart of every experiment conceived, with the highest precision possible. As shown in Fig.~\ref{fig:qcdish}, nowadays the calculation of observables using perturbative and non-perturbative QCD-inspired tools or QCD-inspired effective theories and their connections, makes it possible to do Standard Model precise calculations and to look for New Physics (see for example~\cite{higgstools}). QCD in heavy-ion physics has flourished (via many effective theories and extensions to QFT at finite temperature) that provide new tools to describe the initial conditions of ions moving at relativistic velocities, the intermediate stages where the QGP has formed and the later stages where the high temperature and density of gluons and quarks, become hadrons and jets (see for example~\cite{iancu}).

\subsection{The running and confinement}

\begin{figure}
\centering\includegraphics[scale=0.45]{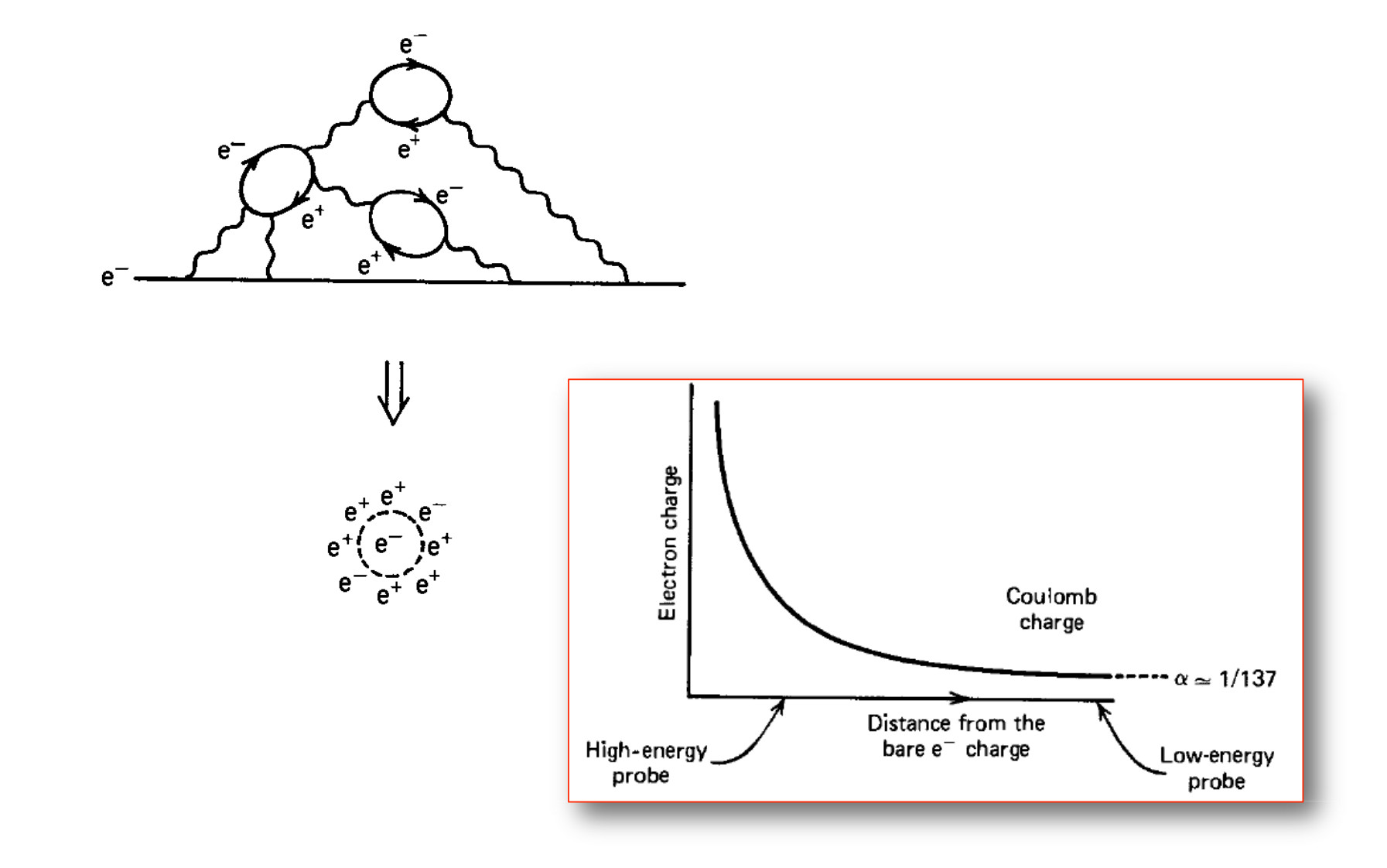} ~\includegraphics[scale=0.55]{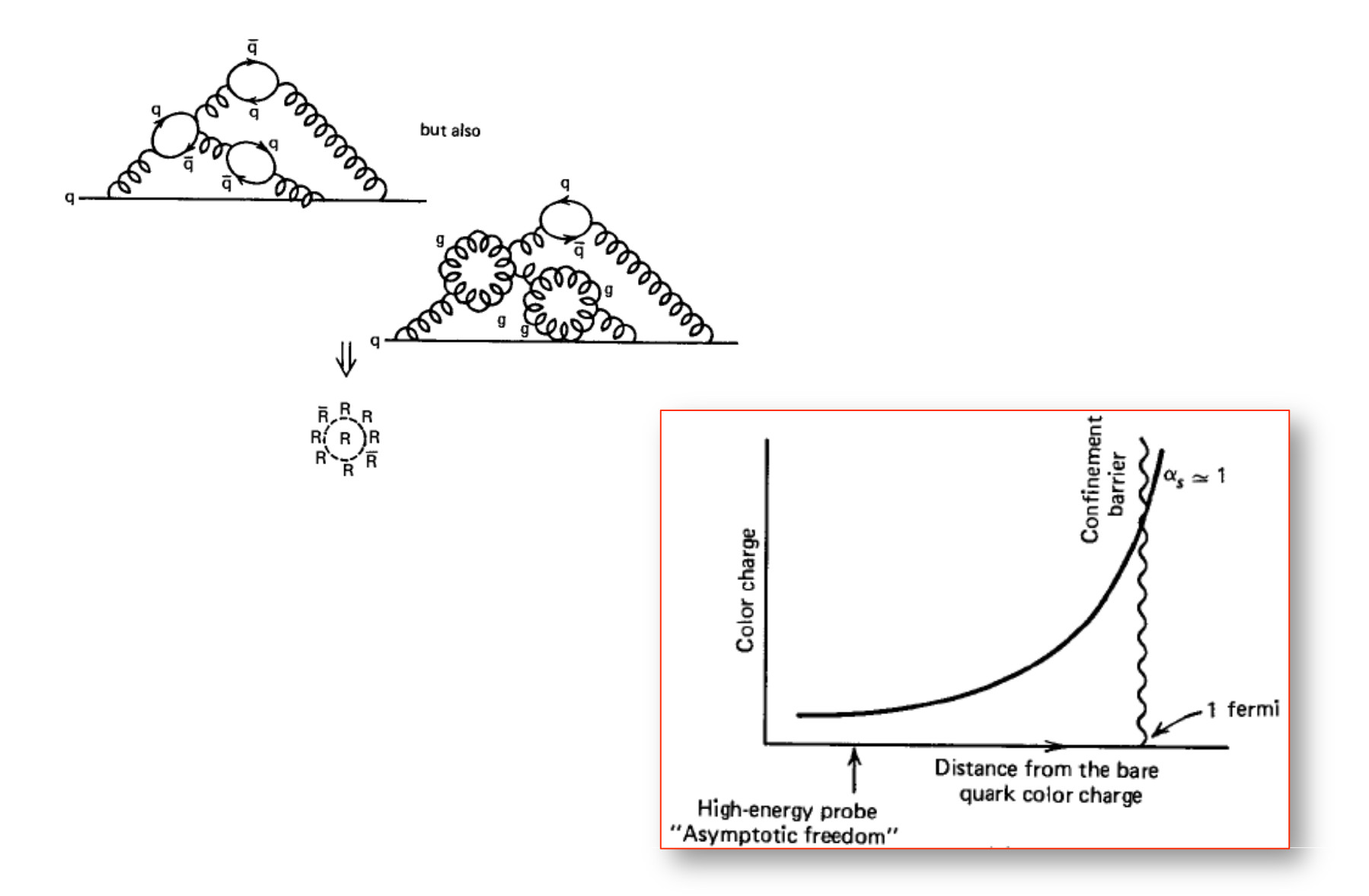}
\caption{On the left we can see the screening effect that quark loops provide for the electric charge, which corresponds to the notion of an effective charge $e(r)$ that becomes smaller with larger distances. This is encoded in the QED beta function $\beta(r) = -\frac{de(r)}{d \log (r)}$, which is positive. On the right we see that the color charge gets screening from the quark loops but antiscreening from the gluon loops, which corresponds to the two contributions with opposite signs to $b_0$ in the QCD beta function. The QCD beta function is negative and so the effective strong coupling becomes small at short distances~\cite{alan}.}
\label{fig:screen}
\end{figure}

Under \textit{ordinary} circumstances, quarks are confined within hadrons. The color potential between quarks inside hadrons at long distances is of order 1 fm and is linear. Separation of quarks requires an infinite amount of energy, so this potential makes hadrons combine into zero net color charge hadrons. From the perturbative regime, confinement can be seen as a direct consequence of the gluon self-interaction. Given that the strength of the strong interaction is characterized by the coupling constant $\alpha_s= \frac{g_s^2}{4\pi}$, we can expose its running (made evident by experiment), by demanding that all relevant observables $\mathcal{O}(Q^2;\alpha_s)$ constructed with QCD should be renormalization scale independent. This is encoded in the Renormalization Group Equation as
\begin{equation}
\left[\mu^2\frac{\partial}{\partial\mu^2}+\beta(\alpha_s)\right]~\mathcal{O}\left(\alpha_s,\frac{Q^2}{\mu^2}\right)=0\nonumber
\end{equation}
where the \textit{beta function} is defined as
\begin{eqnarray}
\beta(\alpha_s) \equiv \frac{d\alpha_s}{d\log\frac{Q^2}{\mu^2}}.
\label{betabis}
\end{eqnarray}
The beta function can be obtained perturbatively to study the running of alpha in the high energy regime (small values of  the coupling). At lowest order, 
\begin{equation}
\beta(\alpha_s)=-b_0\alpha_s^2 ~~~~~~\mbox{with}~~~~~~ b_0=\frac{1}{2\pi}\left(\frac{11}{6}N_c-\frac{1}{3}N_f\right),
\label{beta}
\end{equation}
where $N_c=3$ and $N_f$ are the number of color and flavour degrees of freedom of the theory, respectively. We can now solve Eq.~\ref{betabis} and ~\ref{beta}, assuming $b_0>0$ (which is true provided $N_f < 16$) and get the famous \textit{running} of $\alpha_s$:
\begin{equation}
\alpha_s(\mu^2)=\frac{1}{b_0\log(\mu^2/\Lambda^2)}.\label{e88}
\end{equation}
The parameter $\Lambda$ describes the boundary condition of the first order differential equation defining the running of $\alpha_s$, and corresponds to the scale at which the coupling becomes infinity. Since $\alpha_s$ is not an observable, it can contain all the terms that are $\mu$ dependent, in order to achieve a $\mu$-independent observable $\mathcal{O}$ that has a power-series representation in terms of $\alpha_s$. As shown in Fig.~\ref{fig:screen},  this behaviour shows the difference between the screening and anti-screening efects of the color charge as compared to the electric charge. The observables that can be described where perturbation theory can be applied are usually those where the transferred momentum satisfies $Q^2\gtrsim 1$ GeV$^2$ which marks the separation between the perturbative and non-perturbative regimes.

With the aid of Eqs.~\ref{betabis} and \ref{beta}, we can define a transferred momentum value $\Lambda_{QCD}$ small enough such that the coupling blows up
\begin{equation}
1+b_0\alpha_s(\mu^2)\log(\Lambda_{QCD}^2/\mu^2)=0 ~~~~~\Rightarrow ~~~~~
   \Lambda_{QCD}^2=\mu^2~e^{-\frac{1}{b_0\alpha_s(\mu^2)}}
\end{equation}
where $\Lambda_{QCD}$ is a renormalization scheme dependent quantity. In the $\overline{\mbox{MS}}$ scheme and for three active flavors, its value is of order $\Lambda_{QCD}\sim 200 - 300$ MeV. This has a huge impact on the phenomenological applications of pQCD, since all dimensionful QCD results with small transferred momentum, scale with $\Lambda_{\mbox{QCD}}$. In this context, we are tempted to say that the existence of this scale is a key ingredient for the emergence of baryon masses and thus of the mass of the visible universe, and hastely conclude that this is what explains hadron confinement. Nevertheless, the emergence of confinement is much more than this, so we must pursue an all ecompasing strategy to understand confinement and the breaking of chiral symmetry in QCD, which is also connected to the emergence of hadron masses.

\subsection{Chiral symmetry in QCD}
\label{sec:chiral}

The QCD lagrangian
\begin{equation}
   {\mathcal{L}}_{QCD} = \sum_{\alpha=1}^{N_c^2-1} \sum_{a,b=1}^{N_c} \sum_{i=1}^{N_f}\overline{\psi}_i^a\left(i\gamma^\mu(\partial_\mu\delta^{ab} 
   + i ~g_s ~A_\mu^{ab}) - m_i\delta^{ab}\right)\psi_i^b
   -\frac{1}{4}G^\alpha_{\mu\nu}G_\alpha^{\mu\nu},
\label{qcdlag}
\end{equation}
is that of a non-abelian gauge theory with a local symmetry group $SU(N_c)$ with $N_c=3$ where the fundamental fields are matter fields $\psi^a_i$ (quarks) with masses $m_i$ and massless gauge fields $A^\alpha_\mu$ (gluons). In this lagrangian the strong coupling $\alpha_s=g_s^2/4\pi$ enters both in the quark-gluon interaction term $\overline{\psi}^a\gamma^\mu A^{ab}_\mu\psi^b$ (with $A_\mu^{ab}=A_\mu^\sigma(\tau_\sigma)^{ab}$) and in the gluon self-interaction term through the field strength tensor, defined as $G^\alpha_{\mu\nu}=\partial_\mu A_\nu^\alpha - \partial_\nu A_\mu^\alpha + g_s~ f^{\alpha\beta\sigma}A^\beta_\mu A^\sigma_\nu$. The $N_f$ quark fields  belong to the fundamental representation of the color group ($N_c$ -dimensional), antiquark fields to the complex conjugate of the fundamental representation (also $N_c$ -dimensional) and gluon fields to the adjoint representation ($N_c^2 - 1$ -dimensional), where $a,b$ run from 1 to $N_c$ and $\alpha,\ \beta,\ \sigma$ run from 1 to $N_c^2 - 1$.

If we imagine a universe in which all the quarks have the same mass $m$, then the quark sector of the QCD lagrangian in Eq.~\ref{qcdlag} is (we focus on the flavour indices and we omit colour indices for the purpose of the discussion)
\begin{equation}
   {\mathcal{L}}_{q}=\sum_{i=1}^{N_f}\overline{\psi}_i\left(i\gamma^\mu(\partial_\mu
   + i ~g_s ~A_\mu) - m\right)\psi_i \nonumber
\end{equation}
This lagrangian is invariant under continous global \textit{vector flavour} transformations of $SU(N_f)$ since $\psi_i \rightarrow \psi_i'=e^{-i\alpha^B(T^B)_i^{\,\,\, j}}\psi_j$. In fact using Noether's theorem, we find $N_f^2 - 1$ conserved currents associated with this symmetry since $\partial^\mu j_\mu^A=0$ for 
\begin{equation}
   j_\mu^A(x)=\overline{\psi}_i(x)\gamma_\mu(T^A)^j_{\,\, i}\psi^i_j(x),
   \label{jmuA}
\end{equation}
where $A,B,C=1,...,N_f^2-1$. The charges (or generators) are obtained from the space integration of the density current $j_0^A$
\begin{eqnarray}
   Q^A=\int d^3x\ j_0^A(x), \nonumber
\end{eqnarray}
where $Q^A$ satisfy the $SU(N_f)$ algebra $[Q^A,Q^B]=if^{ABC}Q^C$. Furthermore, current conservation implies that these charges are constant in time, in the sense that they commute with the Hamiltonian $[H,Q^A]=0$ of the theory. The transformation properties of the fields translate into transformation properties of the states (e.g. $Q^A |{\mathbf{p}},i\rangle = (T^A)^j_{\,\, i}|{\mathbf{p}},j\rangle$, supressing spin indices) and if the vacuum state is also invariant under these transformations, all one-particle states of the fundamental representation multiplet, have equal masses $m$. In quantum mechnics, a symmetry of the hamiltonian of the theory reflects in its energy spectrum, most of the time. In this case we have a flavour symmetry transformation that reflects on a degeneracy on the mass spectrum of the theory. This realization of a symmetry is called \textit{Wigner-Weyl mode}. As we will see in what follows, at energies $1.5 - 3$ GeV there is data showing parity doubling in the hadron spectrum which could be explained through this flavour symmetry in a universe we imagined with quarks that have the same mass.

If we modify the universe slightly to now consider quark flavors with different masses
\begin{equation}
   {\mathcal{L}}_{q}=\sum_{i=1}^{N_f}\overline{\psi}_i\left(i\gamma^\mu(\partial_\mu
   + i ~g_s ~A_\mu) - m_i\right)\psi_i,
   \label{lq}
\end{equation}
the mass term now spoils $SU(N_f)$ invariance, therefore currents are not conserved ($\partial^\mu j_\mu^A \neq 0$) and instead of Eq.~\ref{jmuA}, we find 
\begin{eqnarray}
   \partial^\mu j_\mu^A = -i \sum_{i, j=1}^{N_f}(m_i - m_j)\overline{\psi}_i(T^A)^i_{\,\, j}.
   \nonumber
\end{eqnarray}
We could still get back the flavour symmetry in an approximate manner: we could divide quarks into two groups \textit{light quarks} $u,d,s$ and \textit{heavy quarks} $c,b,t$. The mass difference between each group is large ($\gtrsim 1$ GeV) but, an approximate flavour symmetry could still be realised for light quarks: $SU(2)$ for $u,d$ or  $SU(3)$ for $u,d,s$, whereby $\partial^\mu j_\mu^A \sim 0$.

Quarks can also be transformed by means of unitary transformations that include the $\gamma_5$ matrix which are called \textit{axial flavour} transformations. In infinitesimal form these transformations are $\delta\psi_i=-i~\delta\alpha^A~(T^A)_i^{\,\,\, j}\gamma_5\psi_j$ and the current associated with this symmetry is the axial current
 \begin{eqnarray}
   {j_5}_\mu^A(x)&=&\overline{\psi}_i(x)\gamma_\mu\gamma_5(T^A)^j_{\,\, i}\psi^i_j(x).
\end{eqnarray}
Again, if we consider the universe where the quark masses are equal, then under these transformations the invariance of the lagrangian ${\mathcal{L}}_q$ in Eq.~\ref{lq} is spoiled by the mass as $\delta{\mathcal{L}}_q=2i~m~\delta\alpha^A~\overline{\psi}^i(T^A)_i^{\,\,\, j}\gamma_5\psi_j$. So, contrary to the invariance of the lagrangian under vector flavour transformations which is attained with quarks of the same mass, invariance under axial flavor transformations is spoiled even in a universe with quarks of the same mass. Now, of course, in a universe where quarks are massless, the quark lagrangian is invariant under both the vector flavour and axial flavour transformations, which together form the \textit{chiral transformations}. In this case, both the vector and axial currents are conserved: $\partial^\mu j_\mu^A=0, \partial^\mu j_{5 \mu}^A=0$ with the corresponding charges
\begin{eqnarray}
   Q^A=\int d^3x\ j_0^A(x),\,\,\,\,\,\,\,\,\,\,\,\, Q_5^A=\int d^3x\ j_{5 0}^A(x),
   \nonumber
\end{eqnarray}
where the $Q^A$ and $Q_5^A$ satisfy the commutation relations $\left[Q^A,Q^B\right]=if^{ABC}Q^C$, $\left[Q^A,Q_5^B\right]=if^{ABC}Q_5^C$ and $\left[Q_5^A,Q_5^B\right]=if^{ABC}Q^C$. 

In this context, the axial charges do not form an algebra but, if we define left-handed and right-handed charges as $Q^A_L=\frac{1}{2}(Q^A - Q^A_5)$ and $Q^A_R=\frac{1}{2}(Q^A + Q^A_5)$, the newly defined charges decouple and operate separately $\left[Q^A_L,Q^B_R\right]=0$, generating each an $SU(N_f)$ group as $\left[Q^A_L,Q^B_L\right]=if^{ABC}Q^C_L$ and $\left[Q^A_R,Q^B_R\right]=if^{ABC}Q^C_R$. The chiral group is then decomposed into the direct product of two $SU(N_f)$ groups: $SU(N_f)_L\otimes SU(N_f)_R$.

\begin{figure}
\centering\includegraphics[scale=0.3]{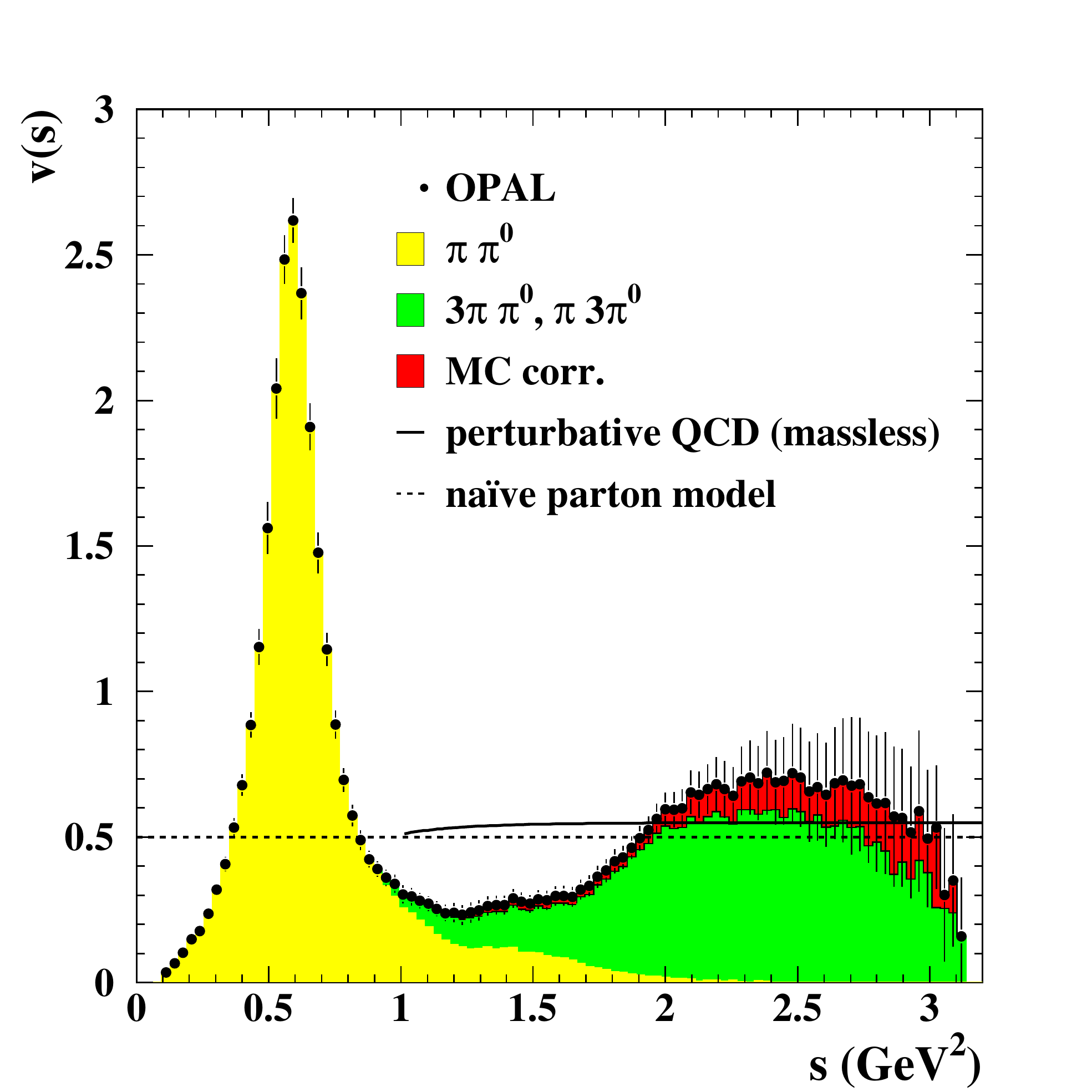}\includegraphics[scale=0.3]{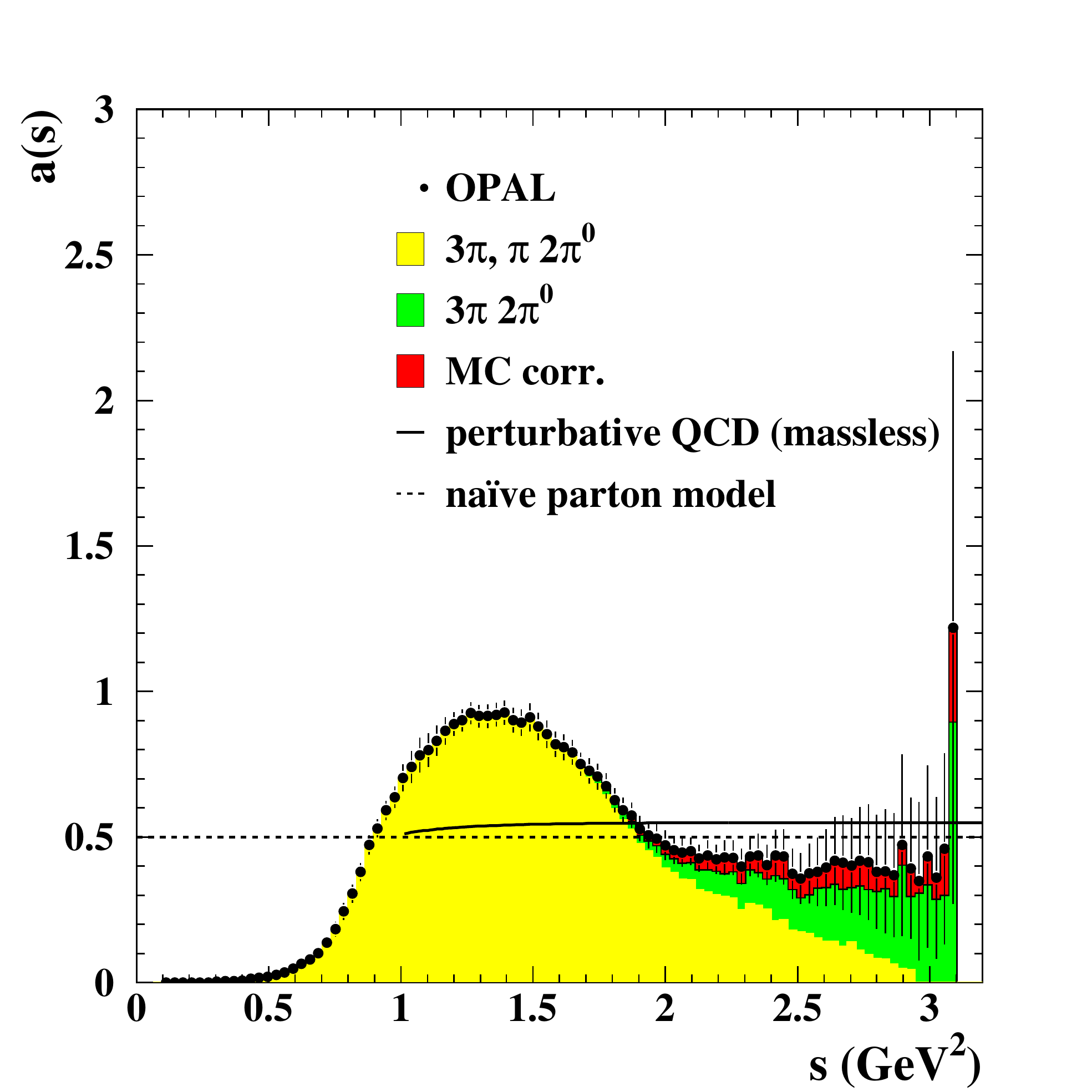}
\caption{OPAL Collaboration measurement of the $\rho$ (vector meson, $I^G/J^P = 1^+/1^-$) and $a_1$ (axial-vector meson, $I^G/J^P) = 1^-/1^+$) mass spectral distributions. The $\rho$ spectrum peaks at $\sim 0.6$ GeV$^2$ while the $a_1$ spectrum peaks at $\sim 1.2$ GeV$^2$~\cite{parityfig}}.
\label{fig:partners}
\end{figure}

It turns out that in our universe, chiral symmetry is not exact, quark masses break it explicitly. In practise, in the light quark sector chiral symmetry is approximate, so the breaking can be treated as a perturbation. How can we \textit{see} that this symmetry is broken? and, what is the signature of this approximate symmetry?. Well, suppose the symmetry is realized in the Wigner-Weyl mode. In the massless quark limit a chiral transformation acting on a massive state changes the parity of the state $Q^A_5 |M,s,{\mathbf{p}},+,i\rangle = (T^A)^j_{\,\, i}|M,s,{\mathbf{p}},-,j\rangle$.
This means that we should find parity partners for the massive states. Even in the massive quark limit, when we consider the light quark sector, the degeneracy within parity doublets is lifted, but the masses should remain close to each other. We do not observe parity doublets (copies of, say the neutron and the proton) in the hadron spectrum, so under these conditions, the Wigner-Weyl mode realization of chiral symmetry does not happen. We could then consider another mechanism, the spontaneous breaking of chiral symmetry, also referred to as \textit{Nambu-Goldstone mode}. In this case the generators do not annihilate the vacuum ($Q^A_5 |0\rangle \neq 0$) and in fact new pseudoscalar states are created: $Q^A_5 |0\rangle = |A,-\rangle$, with $A=1,...,N_f^2 - 1$.

In the massless limit, the charges commute with the hamiltonian therefore these states are $N_f^2 - 1$ pseudoscalar massless particles, Nambu-Goldstone bosons. If $N_f=3$, this would correspond to eight pseudoscalar bosons with small masses. Indeed, the observed pseudoscalar mesons made up of only $u, d, s$ quarks form an octet with $ J^P=0^-$: $\pi^0, \eta, \pi^+, \pi^-, K^0, K^+, K^-, \bar{K^0},$ which are the lightest mesons. Beyond the massless limit, an important measurement involves comparing the experimental mass spectral distributions of parity partner mesons. The main ingredients for these mass spectral distributions are the vector and axial correlators $\mathrm{Im}\Pi_V(q), \mathrm{Im}\Pi_A(q)$. Chiral symmetry implies that these correlators should be identical to all orders in perturbation theory and they were measured at LEP with $\tau$ decays into even and odd number of pions~\cite{paritydoubling-exp}. In particular, the seminal measurement of the $\rho$ (vector meson, $I^G/J^P = 1^+/1^-$) and $a_1$ (axial-vector meson, $I^G/J^P) = 1^-/1^+$) mass spectral distributions are shown in Fig.~\ref{fig:partners}. The $\rho$ spectrum peaks at $\sim 0.6$ GeV$^2$ while the $a_1$ spectrum peaks at $\sim 1.2$ GeV$^2$. This large difference cannot be due to finite current quark masses since, as we already discussed, this produces small differences compared with the masses of the mesons themselves. In this case the differences are of the order of the mass of the $\rho$, so the mechanism that could help explain this, is the spontaneous breaking of chiral symmetry in vaccum\footnote{For further details see for example Ref.~\cite{paritydoubling}}.

\subsection{The many faces/phases of nuclear matter}
\label{sec:faces}

\begin{figure}
\centering\includegraphics[scale=0.25]{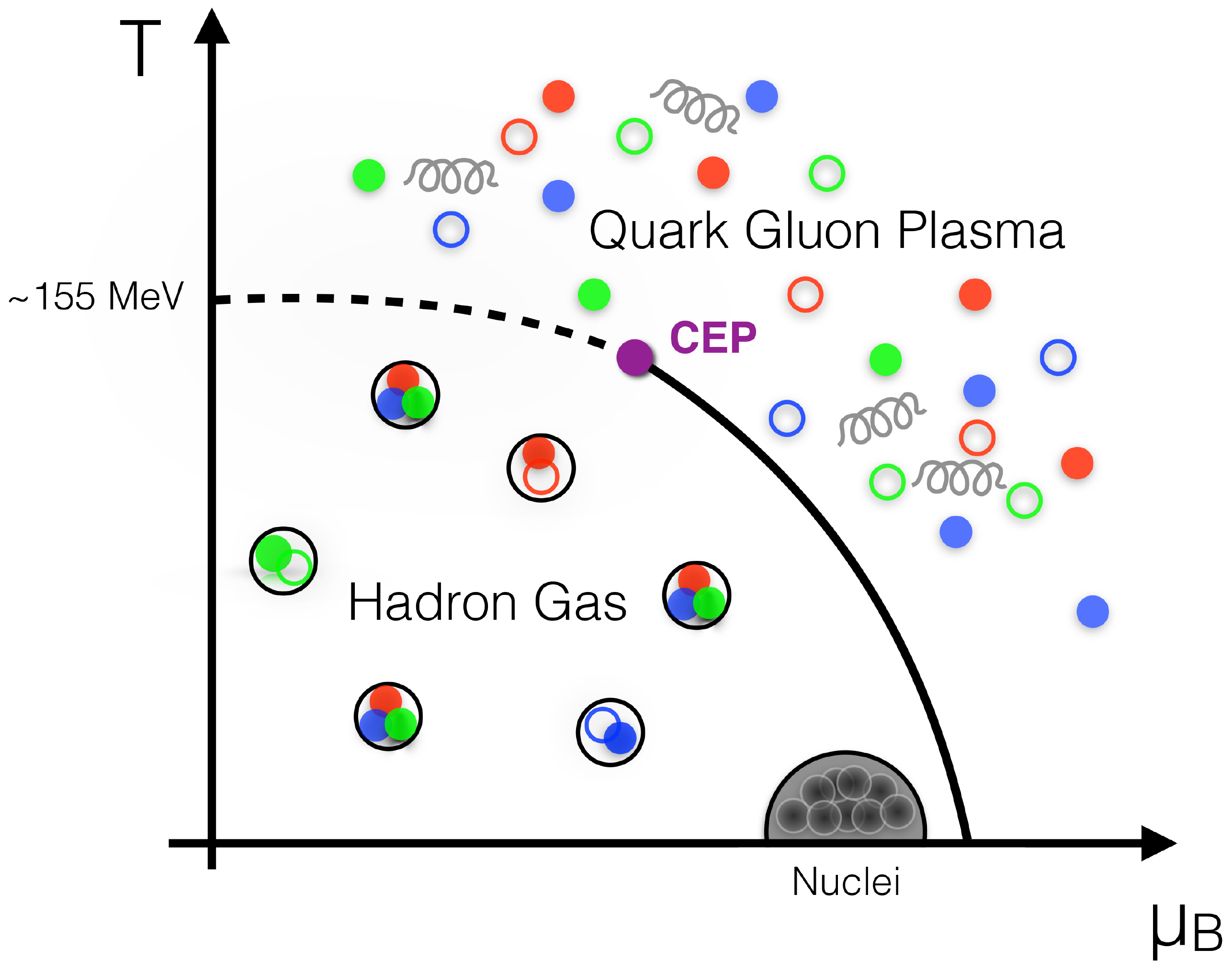}~~~\includegraphics[scale=0.8]{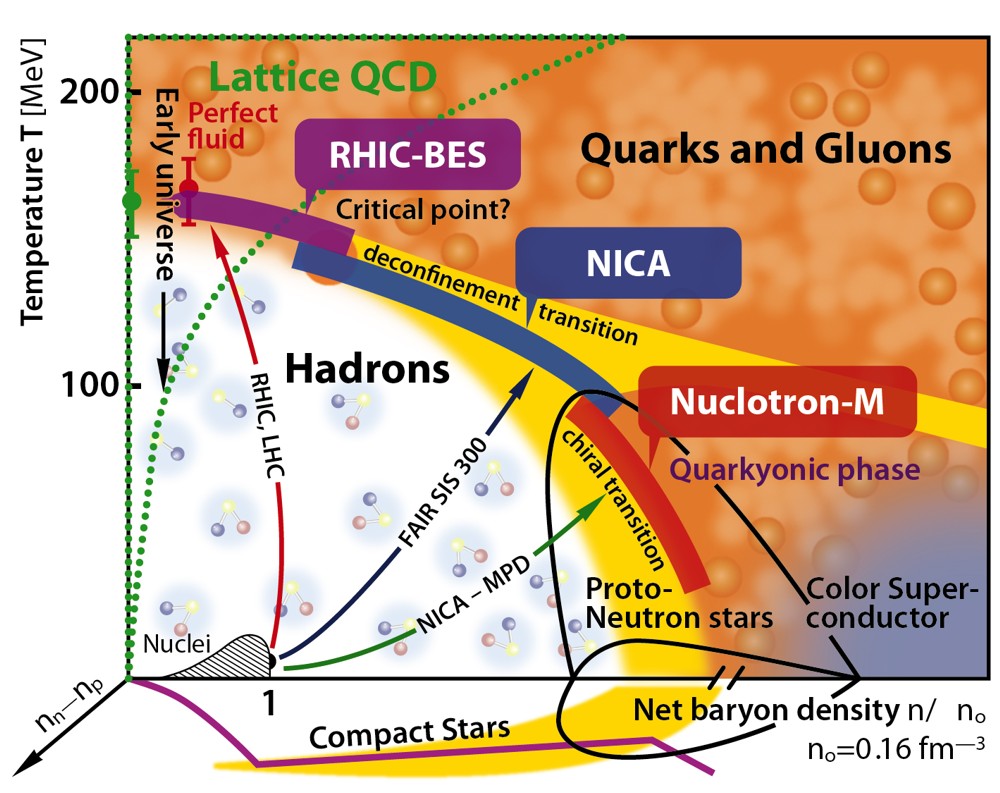}
\caption{On the left, the phase diagram for QCD matter shows the Hadron Gas phase and the Quark-Gluon Plasma phase in the temperature ($T$) and baryon chemical potential ($\mu_B$) plane~\cite{phase-diag-conj}. On the right, the phase diagram for QCD matter shows the Hadron Gas phase, the Quark-Gluon Plasma phase and the Quarkyonic phase in the temperature ($T$) and baryon density ($n$) normalized to the cold nuclei baryon density ($n_o$) plane~\cite{phase-diag-NICA}.}
\label{fig:PhDiag}
\end{figure}

In heavy-ion collisions at relativistic energies, we reach the conditions under which confinement is no more and we obtain a medium of thermally equilibrated hadronic matter where now the quarks and gluons can move freely. In order to describe these conditions, we identify two important parameters: the temperature $T$ and the baryon number density $n_B$ (or its conjugate variable, the baryon chemical potential $\mu_B$). Now, as we discussed before, the intrinsic scale in QCD is $\Lambda_{QCD}\sim 200$ MeV, so we expect a partons-to-hadrons phase transition around $T\simeq\Lambda_{QCD}\sim 200$ MeV and $n_B\sim \Lambda_{QCD}^3\sim 1$ fm$^{-3}$. Also, the coupling constant runs towards smaller values with increasing energy scale so we can anticipate that both confinement dissolves and this may be somewhat connected to the fact that chiral symmetry breaks. We could argue that QCD matter must undergo phase transitions at high baryon and/or energy densities. The anticipation and excitement of the posibility to create and probe these \textit{novel} (back then!) phases of matter was proposed~\cite{faces} just a decade after the birth of the hadron Quark Model~\cite{partons} and two years after QCD was established as a non-Abelian quantum field theory with asymptotic freedom~\cite{octet}\footnote{For more historical insight into the early developments of the QCD phase diagram and the early discussions on the Quark-Gluon Plasma, see for example Ref.~\cite{pheno-review-QGP}.}.

In Fig.~\ref{fig:PhDiag}, the panel on the left is a conjectured phase diagram for QCD matter that shows the Hadron Gas phase and the Quark-Gluon plasma phase separated by a phase transition line that ends in the \textit{critical end point} (CEP). At $\mu_B\sim 0$ the transition is more of a countinous crossover, which as mentioned before, must happen around $\Lambda_{QCD}$. Recently, lattice calculations and effective model theoretical estimations have provided constraints and have allowed for a better picture of where the CEP might be located and for a better determination of the phase transitions and this has been matched by the identification of observables relevant for criticality studies~\cite{locationCEP}. This has been accompanied by several new heavy-ion collision experiments that will be able to scan this phase diagram, beyond what current experiments have achieved\footnote{If you are interested on the thermodynamics and statistical mechanics tools to study QCD and hadrons at high temperature, you can start here~\cite{QGPbooks} and go from there.}. This is schematically shown in the panel of the right in Fig.~\ref{fig:PhDiag} where now three phases are separated by a chiral phase transition, a deconfinement transition and a pseudocritical crossover line. The experimental scans using heavy-ion collisions are shown for the CERN-LHC, RHIC-BES, FAIR-SIS and JINR-NICA programs and where the CEP location can be studied using overlapping scan programs. 
The QGP created at RHIC and LHC at high energies, contains almost equal amounts of matter and antimatter, so the corresponding experimental scan, stays close to the vertical axis: low $\mu_b$ or $n_B$, high $T$. The planned collisions at FAIR and NICA, will allow to create quark-gluon plasma with an excess of matter over antimatter, and so the corresponding experimental scans will be able to explore the bulk of the phase diagram.

In the next sections, we will go deeper into some of these features of the QCD phase diagram. For now, let us turn to heavy-ion physics and collision experiments and discuss the basics of their well-established usefulness to scan this phase diagram.

\section{Relativistic heavy ion collisions}
\label{sec:rhics}

\subsection{What happens when two ultra relativistic nuclei collide?}

\begin{figure}
\centering\includegraphics[scale=0.2]{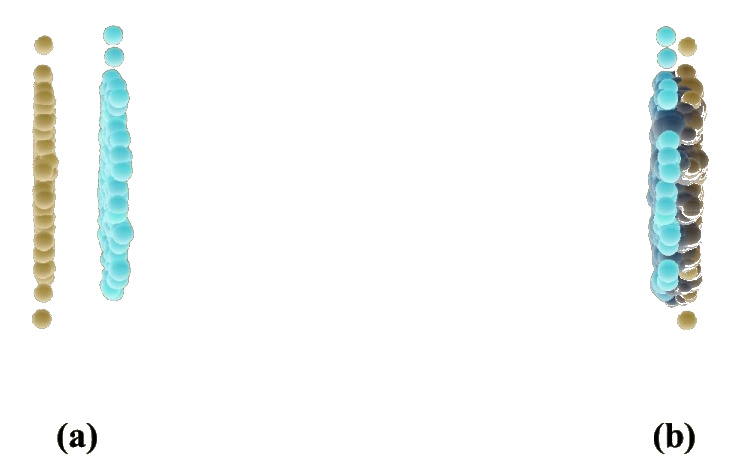}~~~~~~~~~~~~~~~~~~~~~~~~~~~\includegraphics[scale=0.2]{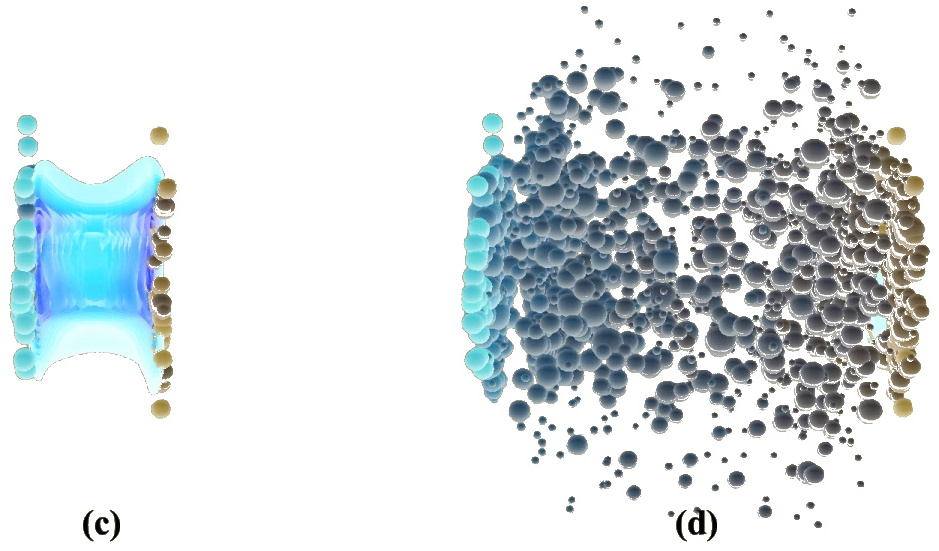}
\caption{Four stages of a heavy-ion collision and its evolution. (a) Shows the two Lorentz contracted heavy-ions moving towards each other in the CM frame, then (b) shows the overlaping nuclei at the collision stage. (c) Just after the collision the heavy-ions travel accross each other and generate a volume of high temperature and energy density and finally in (d), the system expands and cools down, eventually fragmenting into hadrons that travel to the detector.\cite{hiccol}}
\label{fig:HICcol}
\end{figure}

In preparation for a collision, the ions are accelerated to relativistic velocities so much so, that the Lorentz contraction factor $\gamma$ is of the order of 100 at RHIC and of the order of 2500 at LHC. If we choose the laboratory frame as the center-of-mass (CM) frame in a collider, this means that each incident nucleus is a Lorentz contracted disc of the order of $14/\gamma$ fm for Au and Pb nuclei (14 fm being the typical size of these large nuclei). For example, heavy-ion collisions at BNL and CERN have total collision energy ($\sqrt{s_{NN}}$) per nucleon-nucleon pair in the CM frame of 200 GeV at RHIC and 2.76 TeV at LHC. Fig.~\ref{fig:HICcol} shows four stages of the heavy-ion collision and its evolution: (a) first the Lorentz contracted heavy-ions come into the collision stage, (b) then they travel accross each other and generate (c) a volume of high temperature and energy density and finally the system expands and cools down, (d) eventually fragmenting into hadrons that travel to the detector. The stage highlighted in panel (c), is just after the collision where the conditions are optimal for creating QGP. 

The incoming ions are very thin discs of interacting transverse color fields and have on average more quarks than antiquarks with color charges, so that they become sources of colored gluons. When the nuclei pass each other, long color fields fill the space between the receding two Lorentz contracted ions, which makes them loose energy and then they gradually decay into $q-\bar{q}$ pairs and gluons. Now this initial stage is actually dynamic and non-uniform per event, the initial state energy and momentum distributions fluctuate, so modeling the initial state is the first non-trivial task to be approached using high-density/temperature QCD techniques. Furthermore, we know that most of the incident partons lose energy but only a small - albeit important - fraction of them, will undergo hard interactions. These high-$p_T$ hard-interacting partons will be the seed to produce in-medium jets.

On average the energy density around the midpoint between the two Lorentz contracted ions is far grater than the energy density in a hadron (for example at the LHC with $\sqrt{s_{NN}} = 2.76$ TeV, $\langle\epsilon\rangle \sim 12$ GeV/fm$^3 \sim 20 ~\epsilon_{hadron}$). On the other hand, lattice QCD tells us that QCD matter in thermal equilibrium at $T \sim$ 300 MeV has $\epsilon \sim$ $12~ T^4$ = 12.7 GeV/fm$^3$. This means that the medium formed after heavy-ion collisions cannot be described just as a collection of distinct individual hadrons, rather it has to be made out of a high density of quarks and gluons, given the sheer amount of energy density available in a small volume after the collision. Moreover, before the collision the entropy of the two incident nuclei is practicaly null whereas after the collision the final state can contain as many as $10^4$ particles, so this entropy is produced quickly, in the initial moments after the collision.

\begin{figure}
\centering\includegraphics[scale=0.35]{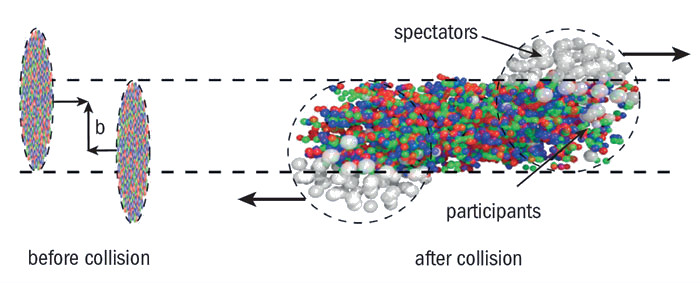}~~~~~~~~~~~~~~\includegraphics[scale=0.4]{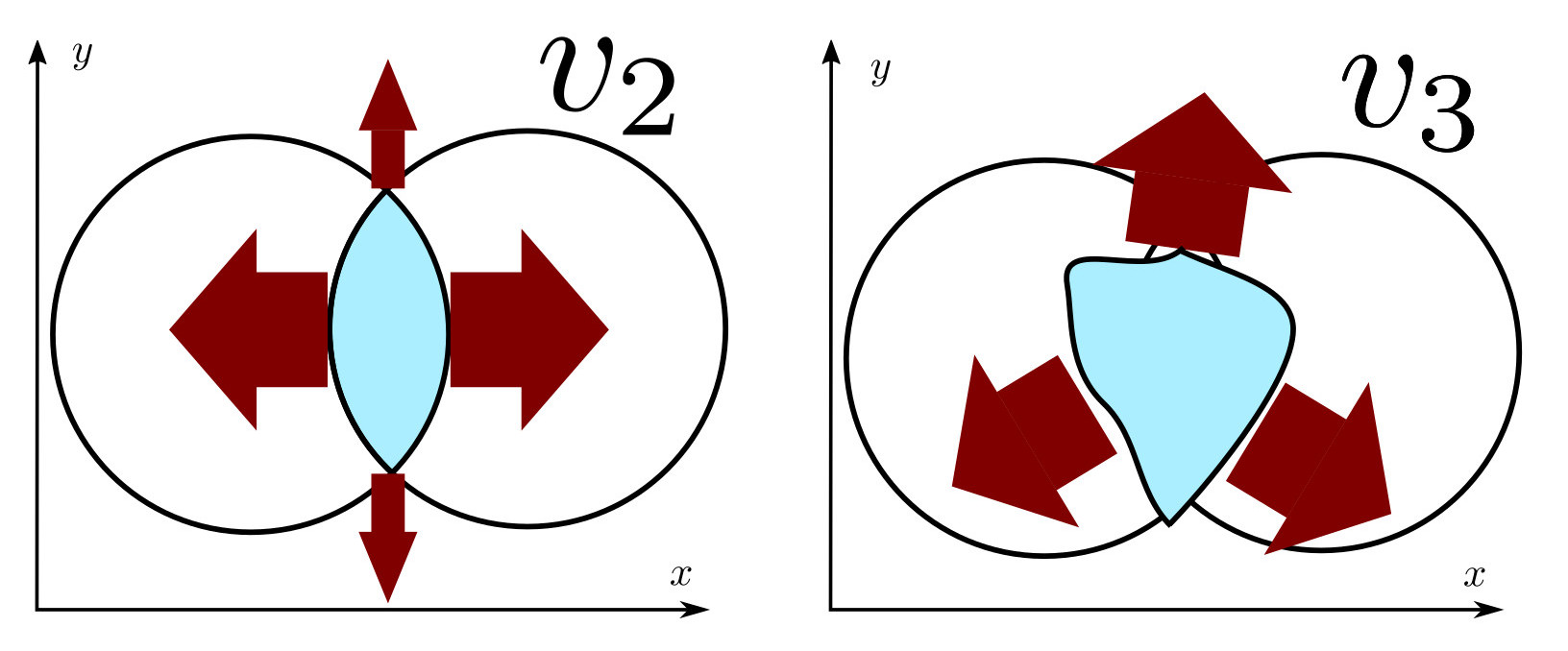}
\caption{On the left, a schematic representation of the collision \textit{centrality}. The ions may collide head-on or may only partially overlap at the collision stage, so in the overlap region there are conditions that facilitate QGP formation. On the right, the collision centrality promotes a characteristic initial shape in the transverse plane to the formed medium, that tends to be more lenticular as the collisions are more head-on or central \cite{centralcern, flowcoeff}.}
\label{fig:centrality}
\end{figure}

Furthermore, as can be seen on the left panel in Fig.~\ref{fig:centrality}, the ions may collide head-on or may only partially overlap at the collision stage, so in the overlap region there are conditions that facilitate QGP formation. This \textit{centrality} of the collision gives also a characteristic initial shape in the transverse plane to the formed medium, that tends to be more lenticular as the collisions are more head-on or central, as shown in the right hand side of Fig.~\ref{fig:centrality}. In fact, the deviations from circular symmetry - or lumpiness - in the initial shape of the QGP and fluctuations of the incident nuclei, lead to preassure anisotropies in the hydrodynamic fluid stage of the plasma. This in turn, leads to anisotropies in the expansion velocity of the products of the hadronized plasma, so that these finally produced hadrons display an azimuthal momentum distribution with the footprint of the initial symmetry or lumpiness. These features can be characterized with the extraction of the \textit{flow coefficients} $v_n$ of the azimuthal particle distribution as
\begin{equation}
\frac{1}{N} \frac{dN}{d\phi} = \frac{1}{2\pi}\left[
   1 + \sum_n 2v_n\cos(n(\phi - \Psi)) \right],
   \label{floweq}
\end{equation}
where $\phi$ is the azimuthal angle and $\Psi$ is the reaction plane, which is determined by the impact parameter and beam direction in a Glauber model, or is determined event-by-event, after the fact, by the weighted projection of the all the $p_T$ tracks.

The extreme limit of an off-center collision, is the case where the nuclei miss each other completly. Still these Lorentz-contracted discs contain charges moving at relativistic speeds, so the electromagnetic fields from each ion do interact. These \textit{ultraperipheral} collisions give rise to $\gamma + \gamma$ and $\gamma + A$ interactions, which dominate the total nucleus-nucleus cross section.

The initial state of the collision can also be characterized using models that allow to correlate the centrality of the collision with the number of \textit{participant} nucleons that collide with at least another nucleon, the number of \textit{spectator} nucleons that do not collide and keep on moving along the beam direction and the number of nucleon pairs that collide or \textit{binary collisions}, assuming transparency of the collision. At early times, simulations show that both spectators and participants create instense but short-lived magnetic fields in the collision zone. The effects of these primordial magnetic fields in the evolution of the collision and the available observables to measure these effects, are a subject of current interest in the community.

\subsection{How do we access the different stages of a heavy-ion collision?}

Eventhough we can use models to simulate the initial condition of a heavy-ion collision as we described before, we only have two quantities under our direct control: which two species of nuclei we accelerate in preparation for the collision and with what amount of energy we make them collide. So with the help of many experiments we have to observe the outcome of the collisions, event by event, and infer the transverse distance between the centers of the colliding nuclei (\textit{impact parameter} $b$) and the location and motion of partons in nucleons and of nucleons in the colliding nuclei. We can select events with a narrow distribution of impact parmeter or \textit{centrality bin} only after the experiment or simulation is done, and we can match this result with the impact parameter distributions expected from the initial condition models. Also, we need nuclear and particle physics studies to support these initial conditions models, e.g. the nuclei can be reasonably well approximated by a collection of nucleons, distributed on average according to three dimensional distributions that provide the usual charge distributions for each type of ion.
Moreover, the average quark and gluon content of the nucleons in nuclei can be given in terms of parton distribution functions (nPDF) which can be sometimes approximated by those of free nucleons (PDF). In fact, the measured hadronic total inelastic cross section can be used to model the nucleons in nuclei as hard spheres with a radius that depends on the collision energy, so that the initial condition of the heavy-ion collision and the subsequent stages can be simulated with well known phenomenological inputs such as $\sigma^{pp}_{inel}(\sqrt{s})$ for $p+p$ collisions.

\begin{figure}
\centering\includegraphics[scale=0.9]{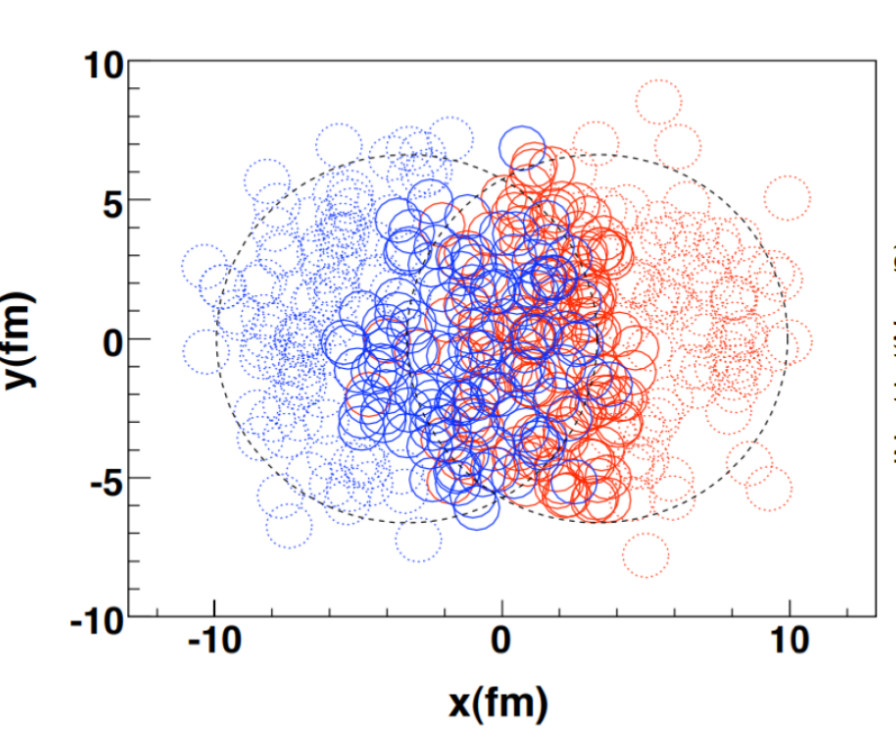}~~~~~\includegraphics[scale=0.35]{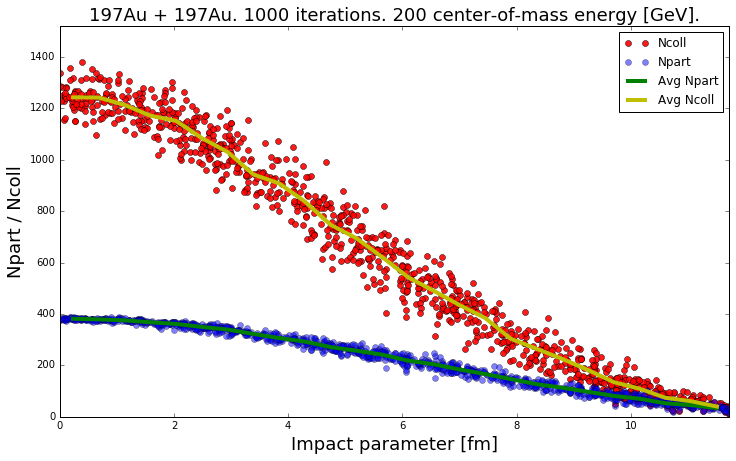}
\caption{On the left, modelling the colliding nuclei as made of transparent spheres where the dashed circles represent spectator nucleons that travel down the beam pipe and the solid circles represent the wounded or participant nucleons which collide with at least one other nucleon. On the right, the number of binary collisions and the number of participants are plotted for 1000 events in Au+Au collisions at 200 GeV with respect to the impact parameter, using MCGlauber\cite{mcglauber}.}
\label{fig:exampleglauber}
\end{figure}

For example, we can model colliding nuclei as made of $A$ transparent spheres of radius $\sqrt{\sigma^{pp}_{inel}/4\pi}$, $A_L$ and $A_R$ being the number of nucleons inside the left- and right-moving nuclei. Then, as shown in Fig.~\ref{fig:exampleglauber}, in every collision, there are $N_{spec}$ spectator nucleons (dashed circles) that travel down the beam pipe, $N_{part}$ wounded or participant nucleons which collide with at least one other nucleon (solid circles), so that $N_{spec} + N_{part} = A_L + A_R$ and $N_{coll}$ is the number of encounters between nucleons in $A_L$ and in $A_R$. So a toy event with a \lq\lq nucleus\rq\rq ~of 8 nucleons lined up in a row that collides head-on ($b=0$) with a \lq\lq nucleus\rq\rq ~of 4 nucleons lined up in a row would have $N_{part} = 12, N_{coll} = 32, N_{spec} = 0$. But in real central heavy-ion collisions, the nucleons at the center of nucleus will hit about 12 nucleons on average, but less if it is located at the edge of the collision. So we expect that in general $N_{coll} \gg N_{part}$, with a more marked difference for the most central collisions. Indeed, this can be seen in the right panel of Fig.~\ref{fig:exampleglauber} where the values of $N_{coll}$ and $N_{part}$ are plotted for 1000 events in Au+Au collisions at 200 GeV with respect to the impact parameter. These results were obtained using the so called \textit{Glauber Model} which is one of the simplest tools that can be used to simulate the initial conditions of heavy-ion collisions, upon which more sophisticated approaches can be constructed where now the Glauber modelling serves as a benchmark.

\subsection{How do we study the quark-gluon plasma in heavy-ion collisions?}

The QGP is made of quarks and gluons that are strongly coupled. They form a collective medium that expands and flows as a relativistic hydrodynamic fluid, with low viscosity to entropy density ratio $\eta/s \gtrsim 1/4\pi$, during the initial stages after the collision $\Delta t \sim 1$ fm/c
(in the fluid rest frame). Just after the Lorentz contracted heavy-ions collide, droplets of QGP fluid form and flow hydrodynamically: the initial high pressure generated by the collision, drives fluid motion, expansion, and subsequent cooling, until the energy density of the droplet is of the order or smaller than that of a hadron. This is when the system hadronizes and a mist of hadrons continues to expand, allowing for hadrons to scatter off each other a few times. Finally these hadrons loose sufficient energy to stop scattering and just stream away freely until they are collected by the detector.

In order to study the quark-gluon plasma in heavy-ion collisions we need adequate experimental probes for the different stages of the formation and evolution of the QGP and a model that incorporates the theoretical ideas relevant for different time and energy scales. Some of the most prominent experimental probes that have provided key information about the different stages of the collision are hadronic probes, such as the relative abundances of pions and kaons, electromagnetic probes such as dilepton and photon production, jet quenching and $J/\Psi$ resonance production and decay~\cite{expprobes}. In the last decade, the nuclear physics community has converged towards an initial consensus on a model with marked epochs in the time evolution of heavy-ion collisions. This model not only helps with the construction of useful physical observables, but it also guides the design of physics plans for current and future experiments. The main features of this model of the stages in the evolution of heavy-ion collisions are as follows~\cite{smallsystem}: \begin{itemize}
\item[I.] The Lorentz-contracted nuclei collide with short transversal time ($\ll$ 1fm/c). The energy deposited into the medium mainly through gluon field interactions, creates and inhomogeneous initial condition in the transverse plane.

\item[II.] At this stage matter is out-of-equilibrium and needs time to equilibrate, so the expansion occurs at almost the speed of light in both the longitudinal direction and radially in the transverse plane.

\item[III.] Now matter is nearly equilibrated and behaves as a fluid (collective modes). We can use viscous relativistic hydrodynamics with and equation of state from lattice QCD to describe the QGP. The small deviations from equilibrium happen because of shear/bulk viscous medium.

\item[IV.] Now, this fluid cools down to about the cross-over temperature of $T \sim 170$ MeV and breaks up into hadrons. This process of hadronization of fluid modes can be modelled using (e.g.) Cooper-Frye hadronization.

\item[V.] Finally, the hadrons scatter inelastically until they reach a \textit{chemical freeze-out}, where no more decays or secondary production is possible and continue to scatter elastically until \textit{kinetic freeze-out}, where their momentum distribution is set. This then produces hadrons in a final-state with momenta as measured experimentally.
\end{itemize}
This model of particle production in heavy-ion collisions with an intermediate epoch during which a hydrodynamic fluid forms and expands, is quite different from the ones used for particle production in elementary hadron collisions in which only a few new particles are created. In fact, the question of whether we can create dropplets of QGP in $p+p$ collisions has caused a big excitement and the posibility to have collective effects in relativistic collisions of small systems is now being pursued with both theoretical and experimental approaches\footnote{For a recent review on small system collectivity effects in relativistic nuclear and hadron collisions see Ref.\cite{smallsystem}.}.

\section{The quark-gluon plasma: extreme QCD} \label{sec:theqgp}

One way to start exploring the basic properties of the QGP is to observe the hadron mass spectra. At a first glance, we can see that there is a \lq\lq QCD mass gap\rq\rq: the pion mass is separated from the rest of the hadron masses. So we could imagine a hadronic phase of nuclear matter (pion gas) where the relevant number of degrees of freedom are 3 ($\pi^-$, $\pi^0$, $\pi^+$) and the rest of the hadrons are basically \lq\lq excited hadron states\rq\rq. Then the corresponding quark-gluon phase (quark-gluon gas) would be one where the relevant degrees of freedom are 37 for two light-quark flavours ($2_{spin}\times 8_{colour}=16$ for gluons, $(7/8)\times 2_{spin} \times 2_{flavour} \times 2_{q\bar{q}} \times 3_{colour} = 21$ for quarks). So the quark-gluon phase has more than ten times the number of degrees of freedom than the hadron phase. If we use basic thermodynamics, we can do back-of-the-envelope estimates of the preassure of a pion gas ${\mathrm p}_\pi$ and compare it with that of a bag of quarks and gluons ${\mathrm p}_{QGP}$. We will find out that
$$ {\mathrm p}_\pi (T) = 3 ~ \frac{\pi^2}{90} T^4 ~~~~~\ll ~~~~~{\mathrm p}_{QGP} (T) = 37~ \frac{\pi^2}{90} T^4 -B,
$$
so even for a bag constant $B^{1/4} \sim 200$ MeV, Nature prefers the system with higher preassure for increasing temperatures ($T > 0.68 B^{1/4}$). So, under these simple arguments a transition from the hadron phase to the deconfined phase is expected~\cite{QGPbookLR}.

\begin{figure}
\centering\includegraphics[scale=0.7]{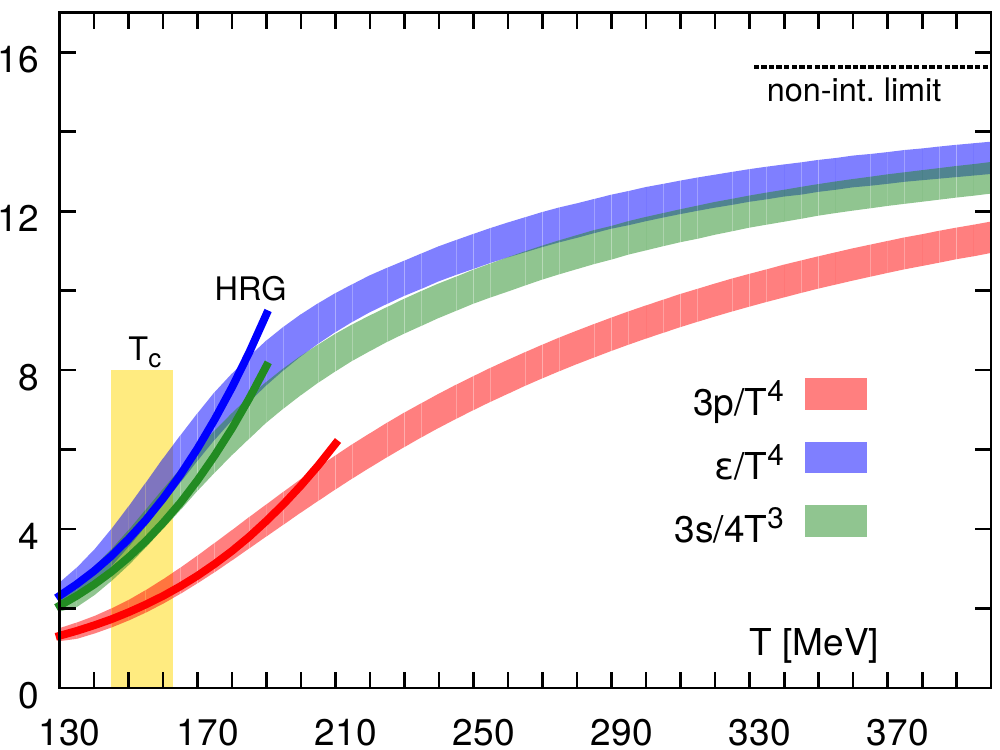} ~~~~~~~~~~ 
\includegraphics[scale=0.4]{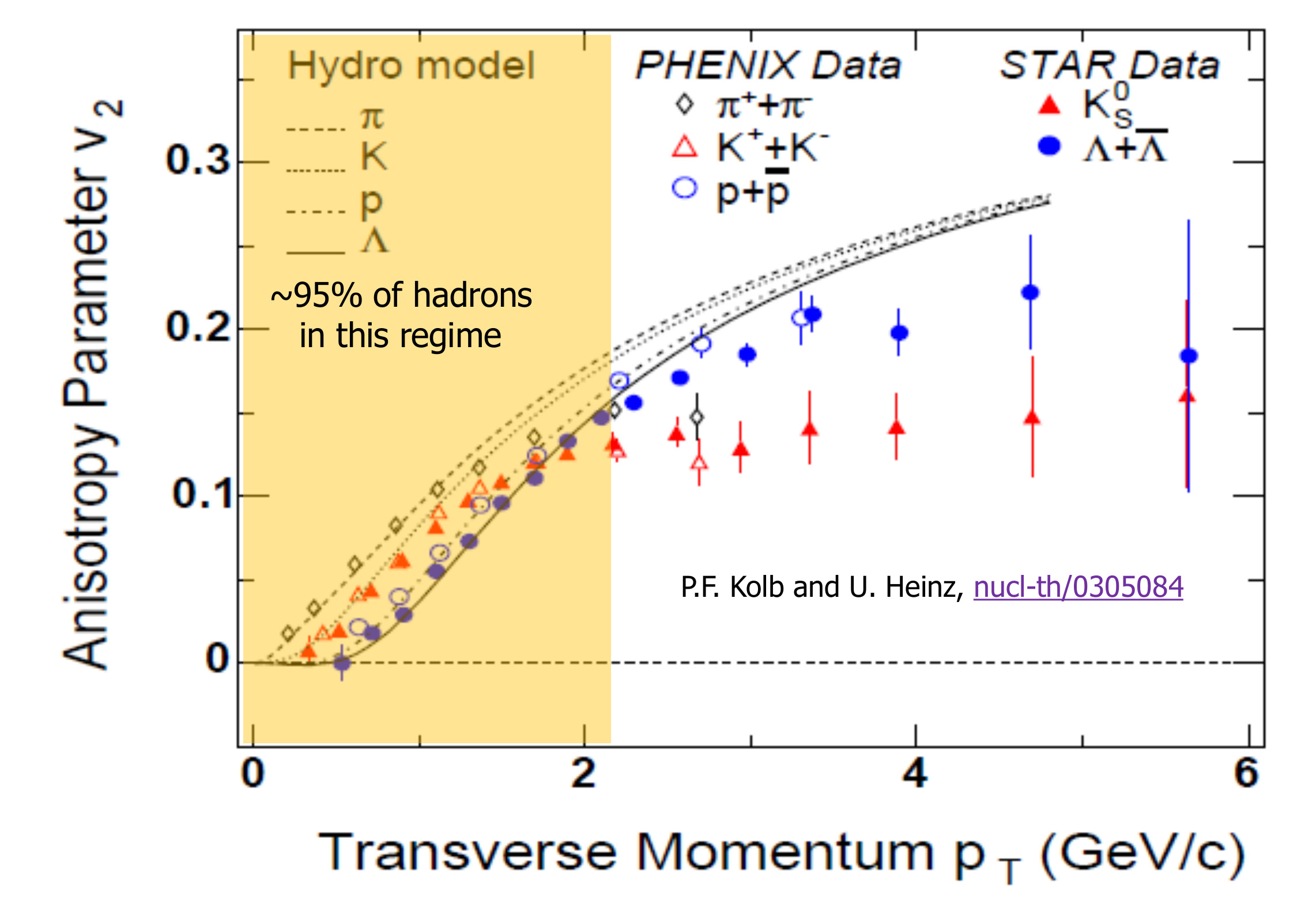}
\caption{On the left, thermodynamic properties calculated with Lattice QCD \cite{hotqcd}, preassure, energy and entropy densities.  On the right, $v_2$ for most hadrons at RHIC which uses a hydrodynamics description of the QGP \cite{Zajc, KolbHeinz}.}
\label{fig:hotqcd}
\end{figure}

Lattice QCD has been able to provide the results not only for the preassure ${\mathrm p}$, but also the energy ${\mathrm \varepsilon}$ and entropy ${\mathrm s}$ densities, for the quark-gluon phase. The left panel on Fig.~\ref{fig:hotqcd} shows these three thermodynamic quantities using solid curves in a temperature range that starts below the critial temperature and goes higher than the temperatures achieved at the LHC. We can see, for example, that the preassure curve (red) is actually $3 {\mathrm p}/T^4$ and the values at both ends of this curve, compare reasonably well with the ones in our back-of-the-envelope estimate, which are for the hadron phase $3\pi^2/30$ and $37\pi^2/30$ for the quark-gluon phase. Also in this figure, it is clear that the values of these thermodynamic properties increase dramatically for temperatures $T \gtrsim 140$ MeV and a smooth crossover is apparent at $T_c = (154 ~\pm ~9)$ MeV. This figure also has the Hadron Resonance Gas (HRG) model curves at low temperatures, which predicts an exponential increase of states. Finally, the non-interacting ideal gas limit is shown at hight temperature which would take an energy density of ${\mathrm \varepsilon}/T^4 = 95\pi^2/60$ to have an ideal non-interacting gas of quarks and gluons. Here we could be tempted to study QGP dynamics with Lattice QCD results above the cross over $T_c$, but this conclusion would be based on a \textit{static} picture, since as we just discussed, thermodynamically this system approaches that of ideal non-interacting gas of quarks and gluons. Now, beyond the static picture, dynamically things look different: the QGP phase is a system more like a viscous hydrodynamic fluid made of strongly interacting quarks and gluons with large cross sections. For example, most of the hadrons produced in heavy-ion collisions at RHIC are best described with a hydrodynamical model of the QGP that expands and generates anisotropies in the transverse momentum distribution, as is shown on the right in Fig.~\ref{fig:hotqcd} for the case of $v_2$ (second coefficient in r.h.s. of Eq.~\ref{floweq}), the so called \textit{elliptic flow}.

One important and seminal work for the applicability of viscous hydrodynamics to the description of the QGP, was reported by Danielewicz and Gyulassy back in 1985~\cite{etaoversbounds}. They came up with a nice way to have bounds for the size of the viscosity in these systems. Since the QGP phase is a system made of strongly interacting quarks and gluons with large cross sections, then we can estimate the viscosity $\eta$ as the transverse momentum in-medium diffusion coefficient, in terms of the number density $n$, average momentum $\langle p \rangle$, mean-free path $\lambda$ and cross section $\sigma$ of the medium constituents
\begin{equation}
\eta \sim \frac{1}{3} n \left\langle p \right\rangle \lambda ~~~~~~~~~~\mbox{or}~~~~~~~~~~ \eta \sim \frac{\left\langle p \right\rangle}{3\sigma},
\label{justeta}
\end{equation}
where on the right hand side, we have taken the mean-free path to be $\lambda = (n\sigma)^{-1}$, so that we can start to see the immediate implications on the small size of the viscosity for large cross sections and low average momentum. If we consider the QGP regime achieved with large chemical potential and/or large temperature, plus the fact that quarks and gluons are almost free under these conditions, then naively we expect small particle correlations, so that the mean-free path is small compared with the system size. 

In order to give a better estimate of the viscosity, we can build a quantity that provides a comparison between the viscosity as a \textit{drag force} and the \textit{inertial mass} of the volume element of fluid, say
$$ \frac{\mbox{drag force}}{\mbox{inertial mass}} \longrightarrow \frac{\eta}{\mathrm{\varepsilon} + \mathrm{p}} = \left(\frac{\eta}{s}\right) \frac{1}{T}  $$
where the energy and preassure densities can be related thermodynamically to the entropy density $s$ as $\mathrm{\varepsilon} + \mathrm{p} = s~T$ at null chemical potential $\mu = 0$. Now, we can use the first expression in Eq.~\ref{justeta} to give a lower bound of $\eta/s$: for a system made out of massless non interacting quanta we know that $s = 4n$ and we can estimate $\left\langle p \right\rangle \lambda \gtrsim \hbar$ so finally the lower bound is
$ \frac{\eta}{s} \gtrsim \frac{\hbar}{4 \times 3}$. Later, using String Theory, Kovtun and collaborators~\cite{stringbound} obtained the bound
$ \frac{\eta}{s} \gtrsim \frac{\hbar}{4 \pi}$.

\begin{figure}
\centering\includegraphics[scale=0.28]{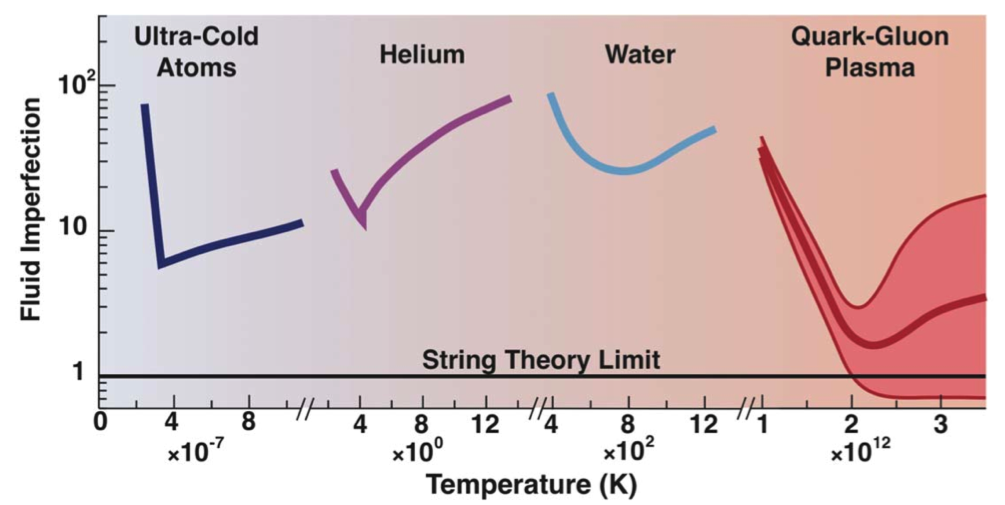} \\
\centering\includegraphics[scale=0.8]{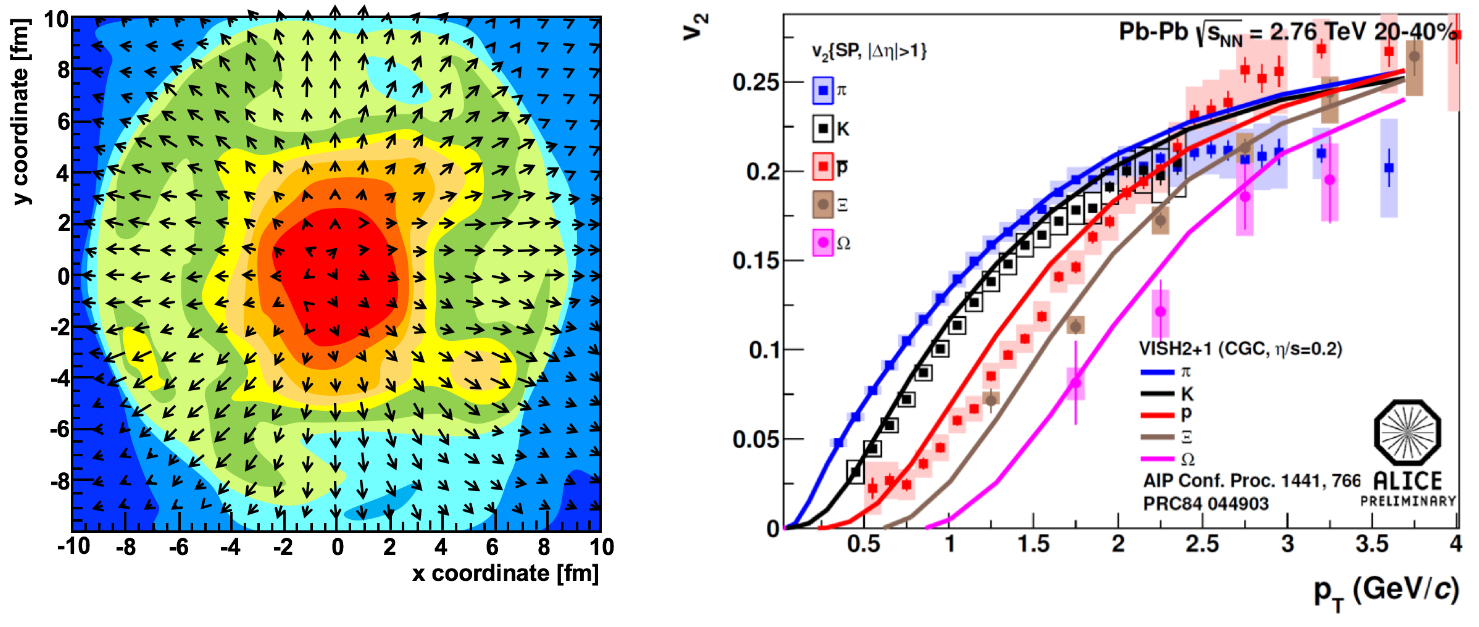}
\caption{In the upper panel, comparison of the ratio $4\pi ~k~\eta/\hbar s$ in natural units~\cite{longrangeplan}. On the lower panels, a viscous hydrodynamics calculation for  a collision between heavy ions, as a one-time snapshot at $t= 5$ fm/c ~\cite{smallsystem,bigquestions}, where the colour scheme indicates the temperature and the arrows indicate the fluid velocity.}
\label{fig:perfecto}
\end{figure}

Soon after, the QGP was popularized as the fluid that might have the smallest viscosity or \textit{fluid imperfection} - a measurement of the departure from ideal hydrodynamics -  with the ratio $4\pi ~k~\eta/\hbar s$. In natural units then the QGP would have near-fluid perfection in the lower bound for $\eta/s$, as is shown in the upper panel of Fig.~\ref{fig:perfecto} in comparison with ultra-cold atoms, helium and water. So, the extreme conditions that prevail in heavy-ion collisions with extreme expansion rates for the fire ball and  
the smallness of $\eta/s$ warrants an extended period of applicability of relativistic viscous hydrodynamics (vHydro) between the pre-equilibrium and the final decoupling epochs. This has resulted in an impresive success in the use of vHydro to describe and predict central observables that help characterize the QGP. For example, the lower panel of Fig.~\ref{fig:perfecto} shows the elliptic flow coefficient $v_2$ for identified hadrons in Pb+Pb collisions at the LHC, compared to hydrodynamic calculations for $\eta/s = 0.2$. The strategy used to obtain the curves shown in this figure, is the numerical solution of coupled relativistic hydrodynamic equations, together with a hadronization and flow analysis as was described in the previous section. There are several very successful public and private codes that solve 2+1 and 3+1 vHydro that can be coupled to codes that generate heavy-ion initial conditions and to codes that help in the later stages of the evolution (freeze-out and hadronization). Even a simple linearized version of vHydro equations  
\begin{equation}
\partial_\mu \delta T^{\mu\nu} = J^\nu ; ~~~~~~\partial_\mu T_0^{\mu \nu} = 0,
\label{lhydro}
\end{equation}
where the medium total energy-momentum $T^{\mu\nu} = T_0^{\mu\nu} + \delta T^{\mu\nu}$ with $T_0^{\mu\nu}$ the energy-momentum for underlying medium in equilibrium and $\delta T^{\mu\nu}$ a \textit{small} perturbation ($J^\nu$ being the source of disturbance), can be solved analytically for certain source models and has provided physical insight into the effects of hydro-modes in experimental observables~\cite{ushydro}.

From the point of view of a microscopic theory, the viscosity is a transport coefficient that quantifies the inefficiency of dissipation in the medium, which would translate into a presistent anisotropy in the energy-momentum tensor $T_{\mu\nu}$. Then in the perturbative picture, the QGP would be made up of quasipartice excitations with momenta the size of the dominant scale, the temperature of the system. So the anisotropy in the energy-momentum tensor, arises from the momentum distributions of these quasiparticles. Then, the scattering and spliting of these constituents will drive the system into its equilibrium value. So in this perturbative approach the challenge is to determine the form of the source of anisotropies by the calculation of the collision operator that enables this system relaxation. At fixed order in perturbation theory, this implies solving a Boltzmann equation using finite temperature techniques
\begin{equation}
 ( \partial_t + \hat{\mathbf{p}}\cdot\nabla_{\mathbf{x}} ) 
f(\mathbf{x},\mathbf{p},t) = - C^{2\to 2}[f] - C^{1\to 2}[f],
\label{boltzmann}
\end{equation}
where $C^{2\to 2}_s$ is a scattering operator, $C^{1\to 2}_s$ is a splitting operator and $f(\mathbf{x},\mathbf{p},t) = dN/d^3\mathbf{x}d^3\mathbf{p}$ is the phase space distribution for either quarks, antiquarks or gluons. The solution of this equation then, is used to obtain the transport coefficients, via the vHydro equations~\cite{etaqcd}. So a full hydrodynamical study, will contain the non-linearized version of Eq.~\ref{lhydro} coupled to a Boltzmann equation simmilar to Eq.~\ref{boltzmann}\footnote{For further details on the applicability of relativistic viscous hydrodynamics and for a complete description of the implementation of vHydro beyond the linearized regime, see for example~\cite{introhydro}.}.


\section{QCD phase diagram and criticality signatures} 
\label{sec:phasecritical}

\subsection{Phase transition, tradition vs praxis}

\begin{figure}
\centering\includegraphics[scale=0.6]{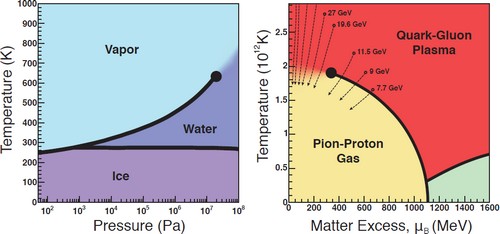} 
\caption{On the left, the phase diagram of water which has continous lines representing the phase transitions and a critical end-point. On the right, the phase diagram of nuclear matter also shows the phase transitions with continous lines and a critical end-point, where the cross-over region begins.~\cite{rajawater}}
\label{fig:waterQGP}
\end{figure}

In general, \textit{phase transitions} refer to the transformations of a substance -from one matter state to another- as a result of variations of external conditions such as preassure, temperature, etc. Typically, when the substance goes through a phase transition, there are certain quantities that often change in a discontinous matter. Perhaps the most familiar phase transitions of a substance, are those of water: in our daily lives we use water in the liquid, solid and gas phases and under certain conditions of, say, preassure and temperature, we know which phase to expect for water. This information is summarized in phase diagrams in Fig.~\ref{fig:waterQGP}, where the solid black lines are the values of temperature and preassure under which the phase transitions occur. Notice there is a \textit{triple point} -three phases coexist- where the phase transition lines intersect and there is a black circle for the \textit{critical end-point} (CEP) where the phase transition line ends. For water, the CEP is known experimentally but, for the QGP, the CEP is predicted theoretically but not known experimentally. The experimental discovery of this CEP is part of current and future heavy-ion collision experiments and the theoretical connection between regions of criticality on a phase diagram and observables at the experiment, are a subject of much interest.

Since phase transitions are the result of interactions of a large number of particles -so size of the system and number of particles are relevant to the discussion- in technical terms, they occur when the free energy -the energetic balance between changes in internal energy and changes in temperature and entropy- is non-analytic (one of its derivatives diverges) for some values of the relevant thermodynamical variables of the system. So, on the phase transition lines the free energies in both phases coincide. For certain systems such as nuclear matter, it is possible to change the state of the substance, without crossing a phase transition line. In this case, the thermodynamic conditions define a \textit{cross-over}, rather than a phase transition. On the right panel of Fig.~\ref{fig:waterQGP}, there is a cross-over region for nuclear matter that transforms between the hadron has phase and the QGP phase, for low baryon chemical potential and high temperature.

Again, in general, phase transitions are characterized as either discontinous or continous. The former are a result of a discontinuous change in entropy at a fixed temperature (the change in entropy corresponds to \textit{latent heat} $L = T \Delta S$) and typical examples are solid-liquid and liquid-gas transitions at temperatures below the critical temperature. The latter involve a continuous change in entropy, which means there is no latent heat in the process, such as liquid-gas transitions at temperatures above the critical temperature.
In practise, sometimes we do phase transition classification \`a la Ehrenfest: the order of a transition is the order of the lowest derivative of the free energy, which shows a discontinuity. For example in boiling water, a first order phase transition is that of a discontinuity in the density, the derivative of free energy with respect to chemical potential, and would be labeled a discontinous phase transition under the previous criteria. In a second order phase transition for the magnetization of a ferromagnet, the derivative of the free energy with respect to the external field is continuous, but the \textit{susceptibility} (the derivative of the magnetization with respect to the external field) is discontinuous and this would be labeled a continous phase transition with the previous criteria. Finally, the modern classification of phase transitions have a more subtle description of them: the first order phase transitions involve latent heat, so the system absorbs or releases heat at a constant temperature, its phases coexist so only some parts of the system have completed the transition; the second order phase transitions are simply continuous transitions and then, the systems' susceptibilities diverge and the characteristic correlation lengths become large.

\subsection{Hadron spectra and Hagedorn temperature}

In order to study the impact of the nuclear matter phase transitions in the hadron spectra, we can start with a simple model for the hadron gas phase: an ideal gas of identical neutral scalar particles of mass $m$~\footnote{For a review on the thermodynamics and hadron spectra of QCD matter is that of Ref.~\cite{reviewhagedorn}.}. The grand canonical partition function for these neutral scalar particles, contained in a box volume $V$ at temperature $T$, assuming Boltzmann statistics is
$2\pi^2\ln{\mathcal{Z}}(T,V) = VTm^2~K_2(m/T)$.
So for temperatures much greater than the mass of these particles, the energy density of the system grows as $T^4$, the particle density grows as $T^3$ and the average energy per particle grows as $T$. The more energy available to the system, the higher the temperatures accesible to it and the more energetic its constituents. Now, if we allow these hadrons to interact and we allow for resonance formation, we can use an improved model of an ideal gas like before, but now with a probability to form certain resonances,
\begin{equation}
\ln{\mathcal{Z}}(T,V)=\sum_{i=0}\frac{VTm_i^2}{2\pi^2}~\rho(m_i)~K_2(\frac{m_i}{T}),
\label{resonancemodel}
\end{equation}
where we sum over all possible resonances ($i=0$ being the ground state) with weights relative to the ground state, encoded in $\rho(m)$. In the early days of hadron discoveries, R. Hagedorn ~\cite{hagedorn} collected the data on the hadron spectra and concluded the hadron spectra could be well described with this model with $\rho(m)\propto\exp(m/T^H)$ where the \textit{Hagedorn temperature} is $T^H\simeq 0.19~\mbox{GeV}$. This hypothesis of the experimental growth with mass of the number of hadronic resonances is at the core of the phenomenology of particle production as a direct consequence of the phase transition between the QGP and the hadron gas. The amount of data we have accumulated since Hagedorn's time,  means that we can revise this model, use it as a guiding light in the search of possible missing states and explore the effects this would have for hadron flow observables~\cite{jacki}. Now, if we are in a regime where not only do we have a high temperature $T$ but also a finite baryon chemical potential $\mu_B$, then this exponetial growth is in direct competition with a Boltzmann factor and $T^H$ is now a fixed temperature limit $T=\left(1-\mu_B/m\right)T^H$, so in this case, the momentum of the constituents of this hadron gas, does not continue to increase and more species of heavier particles can be produced.

\subsection{Chiral transition and hadronization}

In section~\ref{sec:chiral}, we laid out the ideas about how we should expect a phase transition between a state with light current quarks and state with heavy constituent quarks a \textit{chiral phase transition}. In other words, we can study the QCD phase diagram at finite $T$ and $\mu_B$ from the point of view of chiral symmetry restoration, because quarks turn bare in hot and/or dense energetic nuclear matter. For these studies the typical order parameter to follow in the search of a transition is the \textit{chiral condensate} $\langle\overline{\psi}\psi\rangle$, a quark-antiquark bound state with Bose-Einstein statistics. The easiness of this condensate to form in vacuum is characterized by the value $\langle\overline{\psi}\psi\rangle_0=-(0.24$ GeV)$^3$, which sets the scale for the critical temperature of chiral restoration. Chiral perturbation theory provides a strategy to calculate the value of this condensate at finite -but small- temperature and chemical potential and the result shows that the condensate melts, so there is a set of values of temperature and chemical potential that indicate a chiral phase transition in the QCD phase diagram.

In order to probe this chiral phase transition in experiments, we should keep tabs of the particle abundances and make sure that we have a statistical model that includes variations of these abundances due to a high particle density in the phase transition region\footnote{For a review of the statistical-thermal model of particle production in heavy-ion collisions is Ref.~\cite{statmodelrev}.}. For central collisions where $\mu_B \sim 1/\sqrt{s_{NN}}$ and for current collider based experiments where the highest energies are
reached -such as the LHC and RHIC- the bayon chemical potential associated to the reaction is the smallest. In these experiments there is then an increase of the \textit{degree of transparency} of the colliding nuclear matter: the energy injected to the interaction zone, produces roughly an equal number of particles and antiparticles. New experiments -such as NICA and FAIR- have collisions with lower center-of-mass energies, a smaller degree of transparency to create an interaction zone that is rich in baryons.

\subsection{The critical end-point (CEP)}

\begin{figure}
\centering\includegraphics[scale=0.6]{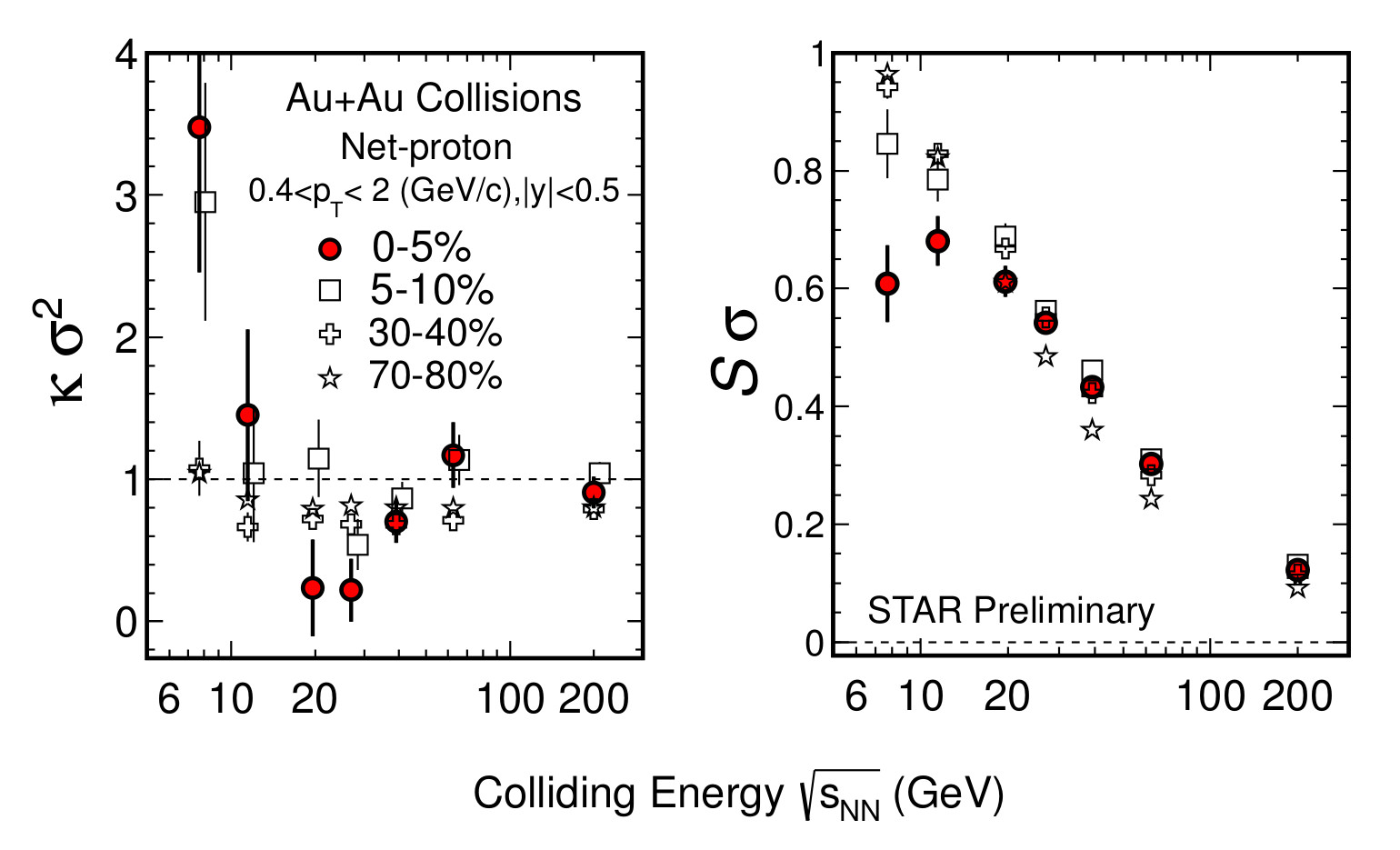}
\caption{Skewness and kurtosis of net-proton number is reported by the STAR Collaboration~\cite{sknetproton}.}
\label{fig:netproton}
\end{figure}

So far, we have complemented the discussion that we started back in section~\ref{sec:faces}, on the theoretical ideas and observed phenomena that motivate a conjectured QCD phase diagram for nuclear matter under extreme conditions of temperature and density, such as the ones in Fig.~\ref{fig:PhDiag}. The main features of the conjectured QCD phase diagram can be summarized as follows: there is a crossover for $\mu_B=0$, there is a first order phase transition that turns into a second order phase transition somewhere in the middle of the phase diagram and there is a CEP somewhere in the middle of the phase diagram, where the crossover becomes a first order phase transition line\footnote{For a review on the location of the CEP in the QCD phase diagram see Ref.~\cite{cep} and references therein.}.

One way to find experimental evidence of the CEP in relativistic heavy-ion collisions, is to look for event-by-event fluctuations of conserved quantities. The possibility to detect \textit{non-Gaussian fluctuations} in conserved charges is one of the tools thought to be sensitive to the early thermal properties of the medium created at heavy-ion collision experiments. In principle, the amount of any conserved charge $Q$ in a given volume of phase space $V$ is an integral over the volume of the charge density $n(x)$, but when the measurement of the charge is performed event-by-event for a volume in a thermodynamical system, we expect thermal fluctuations of $Q$. These fluctuations can be quantified with the help of moments of the charge distribution functions e.g. the variance of $Q$ given in terms of a \textit{correlation function} as 
$$\langle\delta Q^2\rangle_V=\langle(Q-\langle Q\rangle_V)^2\rangle_V = \int_Vdx_1dx_2\langle\delta n(x_1)\delta n(x_2)\rangle.$$

In general for a conserved quantity $c$, the \textit{cumulants} $\langle x^n\rangle_c$ associated to a probability distribution function ${\mathcal{P}}(x)$ of a stochastic variable $x$ are
$$
\langle x^n\rangle_c=\left.\frac{d^n}{d\theta^n}~\ln G(\theta)\right|_{\theta=0},
$$
where $\langle x\rangle_c=\langle x\rangle$, $\langle x^2\rangle_c=\langle x^2\rangle - \langle x\rangle^2 = \langle \delta x^2\rangle$, $\langle x^3\rangle_c=\langle \delta x^3\rangle$, $\langle x^4\rangle_c=\langle \delta x^4\rangle -3\langle \delta x^2\rangle^2$, etc. For a thermodynamical system the partition function $\ln{\mathcal{Z}}$, is the moment generating function - the so called \textit{cumulant generating function}- $\ln G$. Since a conserved quantity will be connected to the volume $V$ of the system in a grand canonical ensemble framework, cumulants are extensive quantities. Notice also that non-Gaussian fluctuations will be driven by non-vanishing higher order cumulants. Since the asymmetry and sharpness of a distribution function can also be established through the \textit{skewness} $S$ and \textit{kurtosis} $\kappa$, then there is a direct connection between these two quantities and the cumulants: when the stochastic variable $x$ is normalized to the square root of the variance $\sigma$, the skewness and the kurtosis are given as the third and fourth-order cummulants: $S=\langle \tilde{x}^3\rangle_c$, $\kappa=\langle \tilde{x}^4\rangle_c$. If we can describe the fluctuations of conserved charges in heavy-ion experiments using hadron degrees of freedom, then the cumulants should be consistent with those of the Hadron Resonance Gas model (HRG) and so deviations from this model are used as experimental signals of the emergence of non-hadron or non-thermal physics. Near the CEP, higher order cumulants change sign and are sensitive to the increase of correlation lengths. In particular when passing through the CEP in a beam energy scan in heavy-ion collision experiments, the fluctuations of multiplicities or mean transverse momenta of particles are a manifestation of higher non-Gaussian cumulants with a non-monotonic behavior of the correlation length. For example, the top panels in Fig.~\ref{fig:netproton} show the skewness and kurtosis for net-proton number as reported by the STAR Collaboration for a beam energy scan from 6 GeV to 200 GeV for Au+Au collisions at RHIC. In peripheral and mid-central collisions, the $\kappa\sigma^2$ values are close to 1 and the $S\sigma$ show strong monotonic increase when the energy decreases, giving the first hints of criticality.

\section{Heavy-ion collision physics: season finale and season premiere}
\label{sec:seasons}

The experiements that have established the production of the QGP -RHIC at BNL and LHC at CERN- and are still providing data to characterise it, represent the golden era of heavy-ion collisions. Now, a new era begins with new experiments -NICA at JINR, FAIR at GSI and RHIC at BNL-,  that are going to provide data to scan regions of the QCD phase diagram that we could not access before. Together with heavy-ion physics, astrophysics and condensed matter will continue to enrich the description of the phases of nuclear matter from different perspectives. Eventhough the collisions are at lower energies, this poses new technological challenges both for the preparation of the ion-beam and for the design of detectors that can have de adequate coverage to capture the products of the QGP evolution. Throughout these lectures, we have shown several examples of the theoretical and experimental efforts that provide the most comprehensive picture, up to today, of the QGP and its evolution. In this section we highlight certain aspects of heavy-ion phenomenology that have resolved some issues in the past, but that also present new challenges in this new experimental era.

\subsection{$R_{AA}$, jet quenching and correlations}

\begin{figure}
\centering\includegraphics[scale=0.4]{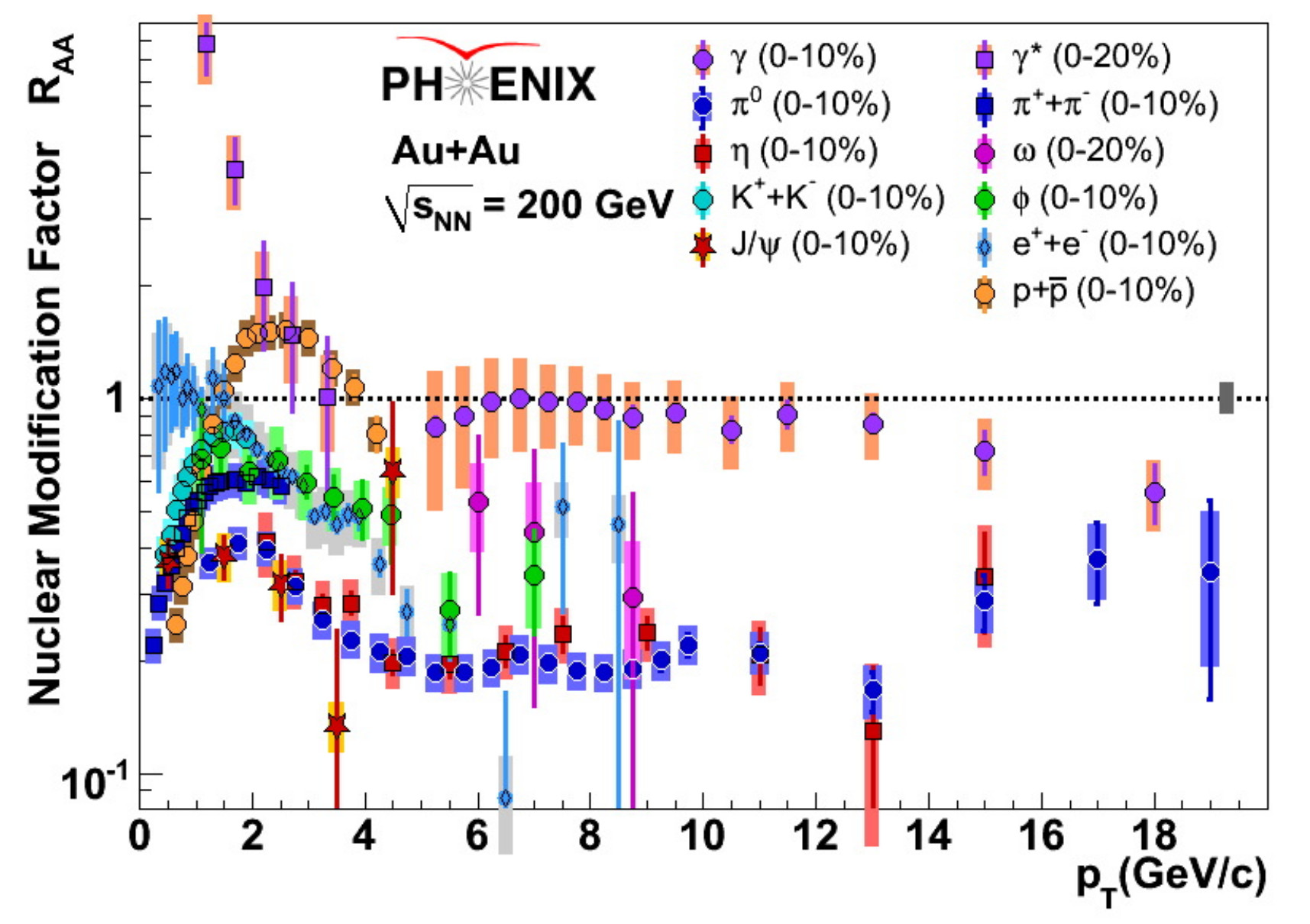} ~~~~~~~~~\includegraphics[scale=0.5]{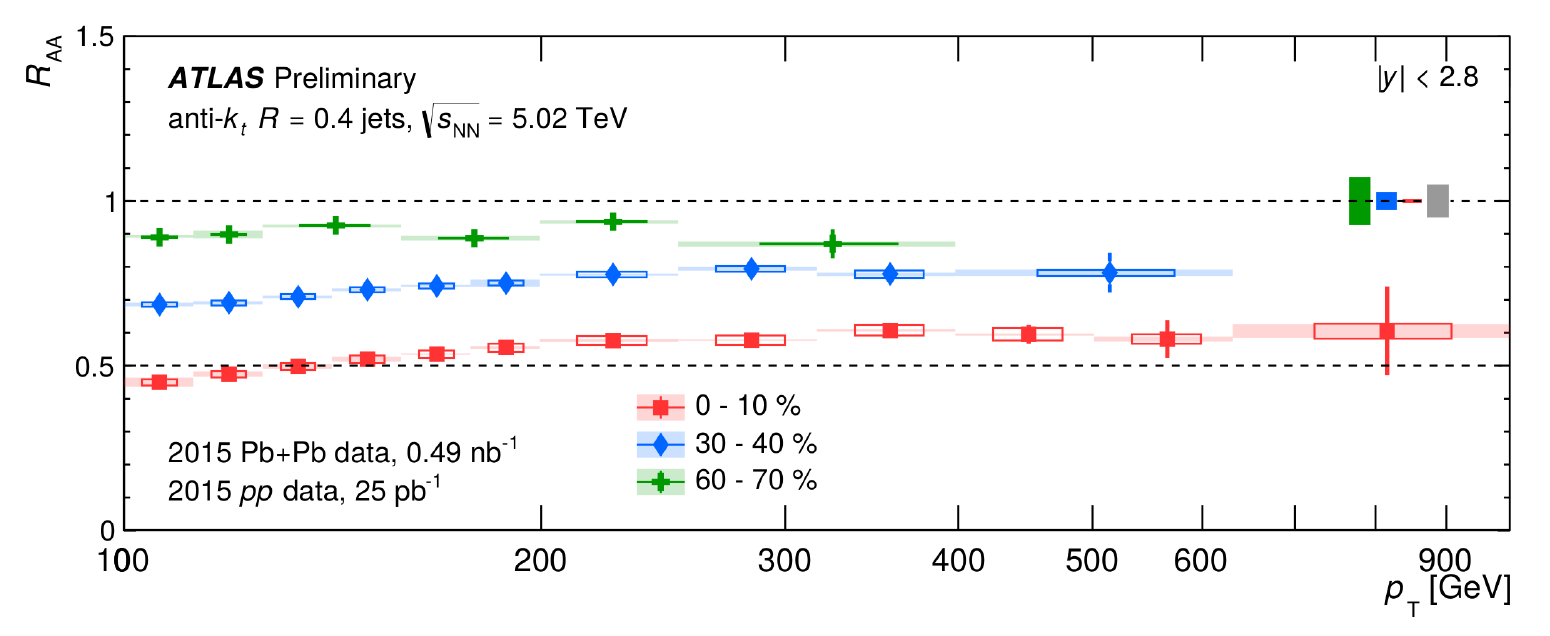} 
\caption{Hadron species show \textit{quenching} at RHIC\cite{phenixquench} and single-inclusive jet quenching at LHC\cite{atlasquench}.}
\label{fig:hadquench}
\end{figure}

In order to quantify the effect of the medium created in heavy-ion collisions in the production of particles, we use the \textit{nuclear modification factor} $R_{AA}$, which is the ratio of the multiplicity of particles produced in $A+A$ collisions with respect to the multiplicity in $p+p$ collisions, normalized to the average number of binary collisions
\begin{equation}
R_{AA}(p_T)=  \frac{1}{\langle N_{\mbox{\tiny{coll}}}\rangle}\frac{\frac{dN^{AA}(p_T)}{dp_T}}
{ \frac{dN^{pp}(p_T)}{dp_T}}. \nonumber
\end{equation}
So, na\"ively, we expect this ratio to be closer to one for primordial photons for which the QGP should be transparent, and perhaps smaller than one for hadrons that form at different stages of QGP evolution. In the left panel of Fig.~\ref{fig:hadquench} we can see the data for $R_{AA}$ for photons and hadrons in the most central $Au+Au$ collisions at $\sqrt{s_{NN}}=200$ GeV, as reported by the PHENIX collaboration~\cite{phenixquench}. Indeed hadrons show large suppression -\textit{quenching}- for a wide range of $p_T$ and photons mostly do not. Looking closely at the plot, we can see that for $p_T <$ 5 GeV there is an excess of photons in Au+Au collisions and there is a slight recovery on the $p_T>$ 5 GeV observed hadron supression. At the LHC this translates into \textit{jet quenching}, a large -by a factor of 2 for most central collisions- suppression
of single-inclusive jets produced in Pb+Pb collisions, when compared to p+p collisions, as shown in the right panel of Fig.~\ref{fig:hadquench}. Jet quenching in heavy-ion collisions is a tool to understanf the formation and evolution of in-medium jets, since there is an enhancement of soft activity at the edges of the jet and far away from the jet. So jet physics in heavy-ion collisions represents one of the most important tools to study both the process of equilibration -through the interactions between the jet and the medium- and gives access to the scale dependence of medium properties. The missing $p_T$ in an event, due to jet quenching, has made it possible to perform in-medium jet-tomography~\cite{missingpt} and to inspire new developments for in-medium fragmentation and jet-formation in Monte Carlo simulations for heavy-ion experiments. Jet angular and rapidity distributions, have shown that there are \textit{ridges} or long-range correlations, that support the emergence of collective phenomena. Since these ridges are present at large rapidities, and show enhacement for less central collisions and lower values of $p_T$, then this complements the expected and observed behavior for the values of $v_n$ and further supports the hydrodynamical modelling of the QGP, that we presented back in Sections \ref{sec:theqgp} and \ref{sec:phasecritical}.

\subsection{Photons as thermometers and viscometers}

Photons are a special probe of the QGP that, unlike the hard-probes, can traverse the medium from which they are emitted without further interaction. The photon spectra provides access to the evolution and temperature of the fireball, since they are emitted at all the evolution stages of a heavy-ion collision. Even more, the mechanisms for photo-production are known from many other experiments, so an excess or suppression of photons produced in A+A collisions, would be a signal of new mechanisms possible due to the conditions that prevail in heavy-ion collisions.

Photons are thermometers of the QGP in the following sense: the effective temperature of the QGP can be extracted from the photon spectrum in the low $p_T$ region, since it represents the inverse logarithmic slope $dN/dp_T \sim \mathrm{Exp}\left\{- p_T/T_{eff}\right\}$. These \textit{thermal photons} provide an estimate of the effective temperature from different experiments, e.g. $T_{eff} = 211 \pm 19$ MeV in 0-20\% Au-Au collisions at $\sqrt{s_{NN}}=200$ GeV at RHIC and $T_{eff} = 304 \pm 51$ MeV in 0-40\% Pb-Pb collisions at $\sqrt{s_{NN}}=2.76$ TeV at LHC. The photon $p_T$ spectrum even shows photons at $T_{eff}<T_c$ with blue-shift to $T_{eff} > Tc$ due to a strong radial hydrodynamic flow for $T_c = 155 - 170$. Recently, there was even a \textit{direct photon's flow \lq\lq puzzle\rq\rq} whereby photons from hard scatterings become dominant at high-$p_T$ and are well described by NLO pQCD plus photons from thermal emission of the QGP and the hadronic phase, but there was a photon excess at low-$p_T$ with large values of elliptic flow, that could not be accounted for by the regular mechanisms~\cite{photonspectra}. This prompted several possibilities for new sources of primordial photons, such as the photo-production via intense magnetic fields present at early stages after the heavy-ion collision~\cite{photonsB}. Photons are viscometers of the QGP in the following sense: using event-by-event simulations of heavy-ion collisions based on viscous relativistic hydrodynamics, we learn that $\eta/s$ suppresses the flow coefficients $v_n$ from the soft photon spectra and the effect is more pronounced for more peripheral collisions~\cite{photonsviscosity}.

\section{Summary and open questions}
\label{sec:final}

In these lectures we presented the basic experimental and theoretical tools to study the collisions - and evolution of the colliding system - in heavy-ion experiments, where the QGP is produced, the \lq\lq simplest form of complex matter described by the fundamental laws of QCD\rq\rq~\cite{bigquestions}. The QGP is probed with a range of experimental set-ups and we can characterize it with several relevant observables. We have a standard model of heavy-ion collisions, that includes the QGP formation and decay, but that also incorporates the formation of hadrons from the primordial fire ball. The QGP is modeled as a relativistic fluid with very small viscosity and this model can describe most of the flow-related observables across a variety of colliding systems and conditions. Finally, the study of phase transitions in nuclear matter using QCD-inspired models, fuels the experimental plans for the discovery of the critial end-point in this diagram and the ultimate understanding of in-medium hadron formation. As was nicely summarized in Ref.~\cite{bigquestions}, the field of heavy-ion physics has spread into new theoretical and experimental areas of development that have provided a framework for the exploration of new \textit{extreme} QCD phenomenology, but that still has many open new and exciting questions to tackle:

\begin{itemize}
\item What new insights can we obtain about well established QCD phenomenology such as hadronization and confinement, from the analysis of heavy ion collision data?
\item Do we have an understanding of all the stages of a heavy ion collision? In particular, how well do we understand the initial stages of the collision process, up to the creation of QGP?
\item Can we create QGP with collisions of small systems such as proton-proton and proton-light ion collisions?
\item What are the properties of QGP?, what have we learned about the dynamics of different probes as they traverse this medium?
\item Which aspects of heavy-ion collisions phenomenology involve weakly coupled dynamics or strongly coupled dynamics? 
\item What are the regions of the nuclear matter phase diagram we can explore using heavy-ion physics?
\end{itemize}

I hope that with these lectures, you are motivated to do hot, dense and exciting QCD, and join the efforts of the heavy-ion community in this era where new experiments are helping us explore the phase diagram of nuclear matter.


\end{document}